\documentclass{article}
\usepackage{amsmath}
\usepackage{amssymb}
\usepackage{braket}
\usepackage[a4paper, left=3cm, right=3cm]{geometry} 
\usepackage{amsfonts}
\usepackage{float}
\numberwithin{equation}{section}
\usepackage{empheq}
\usepackage{xcolor}
\usepackage{lipsum} 
\usepackage[
    backend=biber,
    style=numeric,
    sorting=none,
    doi=false,
    url=true,
    eprint=true
]{biblatex}

\usepackage[
    colorlinks=true,
    citecolor=blue,
    urlcolor=blue,
    linkcolor=blue
]{hyperref}

\begin{document}
\title{Dynamical dark energy from quantum gravity condensates}

\author{
Daniele Granata\thanks{Email: daniele.granata@uniroma1.it}
\\
\small Physics Department, Sapienza University of Rome, Piazzale Aldo Moro 2, 00185 Rome, Italy
\\
\small Istituto Nazionale di Fisica Nucleare (INFN),
Sezione di Roma 1, Piazzale Aldo Moro 2, 00185 Rome, Italy
\\
\small Departamento de Física Teórica \& IPARCOS, Facultad de Ciencias Físicas
\\
\small Universidad Complutense de Madrid, Plaza de las Ciencias 1,
28040 Madrid, Spain, EU
\and
Daniele Oriti\thanks{Email: doriti@ucm.es}
\\
\small Departamento de Física Teórica \& IPARCOS, Facultad de Ciencias Físicas
\\
\small Universidad Complutense de Madrid, Plaza de las Ciencias 1,
28040 Madrid, Spain, EU
}

\date{}

\maketitle

\begin{abstract}
We investigate dynamical dark energy emerging from quantum gravity within the framework of group field theory (GFT) condensate cosmology. We generalise previous models of quantum gravity induced cosmic acceleration by considering a broader parameter space, including an additional scalar matter field besides the relational clock, and accounting simultaneously for the phase dynamics and the contribution of multiple condensate modes. Working in a mean-field approximation with coherent peaked states and restricting to Hermitian GFT interactions, we derive the corresponding effective cosmological dynamics and analyse its late-time behaviour. We show that the system approaches an asymptotic de Sitter regime, dynamically dominated by a single condensate mode, while the approach to this regime can exhibit a non-trivial dynamical dark-energy evolution. In particular, both the condensate phase and subdominant modes can generate deviations from the asymptotic equation of state $w=-1$, including phantom behaviour and, under suitable conditions, a crossing of the phantom divide. The additional scalar matter field enters directly into these conditions, providing a link between the matter content and the quantum gravity induced dark-energy dynamics. These results extend the robustness and generality of the GFT mechanism for dynamical dark energy and provide a basis for confronting the underlying quantum gravity dynamics with cosmological observations.
\end{abstract}

\newpage
\section{Introduction}
A complete understanding of dark energy, or the observed current acceleration of the expansion of the universe, is one of the main outstanding issues in modern cosmology. This is true even if such acceleration is modelled simply by a cosmological constant, as in the standard model of cosmology, the $\Lambda$CDM model, because one still has to establish how quantum fluctuations of matter fields coupled to the gravitational field, and of the gravitational field itself, affect its value, and more generally to what extent its value is determined by microphysics \cite{Carroll:2000fy, Burgess:2013ara}. It is even more of a challenge if dark energy is dynamical, as recent observations seem to suggest \cite{DESI:2025fii, Ozulker2025PhantomCrossing}, as a way to address several cosmological tensions which cast doubts on the $\Lambda$CDM model. Given the importance of the issue, theoretical models of such dynamical dark energy abound \cite{DE-models}. They are usually formulated in the language of field theories of exotic (scalar) matter \cite{quintessence, phantom} or of modified gravity theories \cite{modified}. Such field-theoretic models can be quite successful in accounting for the observations, often at the cost of some intricate and quite ad hoc modelling choices. Still, they appear to be rather phenomenological models, lacking deeper physical foundations and true explanatory power.

\

It is then natural to hope that candidate formalisms for  fundamental physics, i.e. quantum gravity, could provide explanatory mechanisms, or even a full derivation from microphysics, for (dynamical) dark energy.
This may require quantum gravity to go beyond the effective field theory framework altogether, since dark energy affects the large-scale dynamics of the universe, rather than being a small-distance or high-curvature effect. It may thus require quantum gravity to be something more radical than a non-perturbative and background independent quantization of the gravitational and matter fields (already a daunting challenge, of course). 

\

This is a possibility often voiced in the quantum gravity community, where a number of quite radical views are actively explored \cite{Buoninfante:2024yth}.
Indeed, various quantum gravity formalisms propose a picture in which geometry and spacetime itself are emergent notions, resulting from the collective dynamics of novel quantum \lq constituents\rq{} which are themselves not spatiotemporal fields. Spacetime itself would be, in this picture, a sort of peculiar quantum many-body system, and usual spacetime physics in terms of (quantum) fields (including the gravitational one, i.e. spacetime geometry) would be a coarse-grained, macroscopic approximation. This would amount to a deeper sense in which classical spacetime physics would be emergent, with respect to the traditional \lq quantum gravity $=$ quantized GR\rq{} view \cite{Oriti:2018dsg}.    

\

Tensorial group field theories (TGFTs) \cite{TGFTs} propose a possible concrete realization of such an emergent spacetime scenario.
They can be seen as higher-dimensional generalization of random matrix models for 2d gravity \cite{Matrix}, enriched with the type of quantum geometric data identified in canonical loop quantum gravity (and in quantum simplicial geometry). When these additional data endow the TGFT states and amplitudes with a quantum geometric interpretation, the corresponding models have been traditionally referred to simply as group field theories (GFTs), to distinguish them from models with different applications. Indeed, GFTs can also be understood as a second-quantized formulation of the kinematics and dynamics of the quantum states of loop quantum gravity \cite{LQG, GFT-LQG}, recast in the language of quantum many-body physics, with the building blocks of loop quantum gravity states (spin network vertices) playing the role of \lq quanta of space\rq{} and their dynamics expressed in a quantum field theory language \lq outside of any spacetime\rq. The dynamical connection with spin foam models \cite{SF, GFT-SF} and lattice gravity path integrals \cite{latticeQG, GFTlatticeQG}, which appear as Feynman amplitudes of GFT models in their perturbative expansion (and of which GFTs provide a completion, removing the dependence on any individual lattice and implicitly defining a continuum limit \cite{GFT-renorm, TGFT-Landau}) strengthens their appeal as quantum gravity candidates.   

\

As with all other quantum gravity formalisms framed within an emergent spacetime perspective, and thus not expressed in the usual spacetime or effective field theory language, a main challenge for TGFTs is to show that the usual spacetime language and known gravitational dynamics can be recovered in some suitable approximation \cite{Oriti:2018dsg}. 

In TGFTs, thanks to their field-theoretic, albeit peculiar, formulation, this challenge can be tackled using tools from quantum many-body physics, particularly by gaining control over suitable collective states and hydrodynamic-like approximations.
This programme has achieved some remarkable successes in the cosmological context. Indeed, GFT condensate cosmology, in its various implementations \cite{TGFTcosmology} has shown how a semiclassical Friedmann dynamics can be recovered (in quite some generality) \cite{Friedmann} from the underlying quantum gravity dynamics, how the cosmological singularity is naturally resolved by a quantum bounce \cite{bounce}, and how even cosmological perturbations and effective field theory for gravitational and matter fields can be obtained from the fundamental theory \cite{cosmopert,EFTscalar}, among other results.

\

Recently, within the GFT cosmology context, considerable attention has been devoted to the dark energy problem. The general idea being explored is that the fundamental \lq microscopic\rq{} interactions among GFT quanta could produce, at the level of the emergent cosmological dynamics, novel terms driving the acceleration of the universe's expansion. 

This was first shown in a simple phenomenological model \cite{PithisSakellariadou}, in which a single mode of the GFT condensate states was relevant, interactions only depended on the condensate density, and not on its phase, and the GFT quantum dynamics was studied in a mean-field approximation (thus considering only exact coherent states of the GFT field operators). It was found that, under such restrictions, GFT interactions of order six would indeed produce a term corresponding to an effective cosmological constant, in the emergent cosmological dynamics.
More recently, work has gone into generalizing this initial finding, in different directions. In \cite{Oriti:2021rvm,Pang:2025jtk}, again in a simple mean field approximation, the same kind of interactions, but for condensates having two modes excited, rather than just one, produce a cosmological acceleration which has now a non-trivial dynamics itself, thus providing a quantum gravity mechanism for a dynamical dark energy. This was also shown to be, with some generality, of a phantom type, which made the model of even greater interest for cosmological observations. Then, it was shown \cite{Marchetti:2025jze} that the contribution of the condensate phase to the emergent cosmological dynamics can also be important, and indeed represent another, independent mechanism for a dynamical dark energy produced by pure quantum gravity effects. The effective cosmological dynamics obtained in \cite{Marchetti:2025jze} has been also shown to provide a good fit for the most recent cosmological observations, which can be used to constrain the fundamental quantum gravity dynamics, in a remarkable (and quite rare) direct contact between quantum gravity models and observations \cite{LUCA-DESI}. All the above results were obtained in a mean field approximation, and with a cosmological evolution defined in terms of a relational clock (provided an effective scalar degree of freedom) singled out at the quantum level via appropriate choice of \lq coherent peaked states\rq \cite{marchetti_effective_2021} (see also \cite{PWGFT} for another treatment of quantum clocks in a constrained formulation of GFT).

A different route was taken instead in \cite{GFT-DARKSector}, where a \lq deparametrized\rq{}  approach was followed, quantizing the GFT system with respect to a scalar degree of freedom singled out at the classical GFT level. This allowed an analytic derivation of the effective cosmological dynamics from the interacting quantum GFT dynamics, for a broader class of quantum states \cite{Gauss}. Again, interactions of order six, like in the other cases, produce a cosmological acceleration at late times (while interactions of order four contribute a dark matter-like term). 

\

In this contribution, we generalise and then study in its late-time implications the GFT cosmology model of \cite{Marchetti:2025jze}, thus using a mean field approximation and a definition of a relational dynamics via coherent peaked states. Our generalization proceeds along three directions: we consider a slightly larger parameter set, removing some simplifying assumptions adopted in the original construction; we include an additional scalar matter field, which plays the role of a non-trivial matter content of the universe, besides the scalar matter used as a clock; we include in the effective dynamics both the contribution from two condensate modes and from the condensate phase, thus combining the two known mechanisms for producing a dynamical dark energy from quantum gravity interactions. At the same time, since we are focusing on the dark energy problem, we restrict attention to hermitian interactions, contrary to \cite{Marchetti:2025jze}, where non-hermitian interactions have been shown to be promising to model a viable QG-induced inflationary expansion in the early universe.

Our goal is to increase robustness and generality of GFT explanation for (dynamical) dark energy, identified in previous work, so to make also the connection with cosmological observations more robust and compelling.

\section{From fundamental GFT to emergent cosmology}
{\bf The GFT formalism}

Let us introduce briefly the GFT formalism we use in the following. For more details, see \cite{TGFTs}. We consider a GFT field depending on four \lq geometric\rq{} variables and two \lq matter\rq{} degrees of freedom\footnote{The quotation marks serve as a reminder that a proper geometric or matter interpretation for these degrees of freedom should emerge in some appropriate continuum approximation, and here it serves the purpose of guiding model building and interpretation at the discrete level only.}

\begin{equation}
\begin{aligned}
&\varphi: G^4 \times \mathbb{R}^2 \to \mathbb{C},  \\
&\varphi(g_I h, \chi, \eta) = \varphi(g_I, \chi, \eta) \quad \forall h \in G, \, \, I=1,2,3,4\quad .
\end{aligned}
\end{equation}
When looking at the quanta of such field, in a Fock representation, the variables \(g_I\) denote the geometric degrees of freedom associated with the four faces of a quantum tetrahedron, while \(\chi\) and \(\eta\) are discretized scalar matter fields defined on the same tetrahedron. 
The right-invariance condition implements gauge invariance under
rotations of the local internal reference frame associated with
the tetrahedron and enforces a
closure condition. Indeed, for \(G=SU(2)\), which is the choice we use in the following, its classical geometric
counterpart reads $ \sum_{I=1}^{4} X_I = 0$,
where \(X_I\in\mathfrak{su}(2)\simeq\mathbb{R}^3\) are the (flux) normal
vectors associated with the four faces (whose areas correspond to the modulus of the vectors). This discrete geometric interpretation is then confirmed, in GFT models, by the characterization of more general many-body states and perturbative quantum amplitudes.

Upon quantisation, the classical GFT fields are promoted to operators
$\hat{\varphi}$ and $\hat{\varphi}^\dagger$ satisfying the
commutation relations:
\begin{equation}
\begin{aligned}
&[\hat{\varphi}(g_I,\chi,\eta),
  \hat{\varphi}^\dagger(g_I',\chi',\eta')]
 = \mathbb{I}(g_I,g_I')\,
   \delta(\chi-\chi')\delta(\eta-\eta'),\\
&[\hat{\varphi}(g_I,\chi,\eta),
  \hat{\varphi}(g_I',\chi',\eta')]
 = [\hat{\varphi}^\dagger(g_I,\chi,\eta),
    \hat{\varphi}^\dagger(g_I',\chi',\eta')]
 = 0.
\end{aligned}
\end{equation}
Here, $\mathbb{I}(g_I,g_I')= \int_G \mathrm{d}h\,
      \prod_{I=1}^{4}
      \delta_G\!\left(g_I h (g_I')^{-1}\right)$, with $\mathrm{d}h$ the normalized Haar measure on $G$ and $\delta_G$ the delta distribution on the group, is the identity kernel on the
space of gauge-invariant fields.\\
A general GFT action (for more on model building, see \cite{TGFTs, TGFTcosmology,TGFTscalarmatter}) has the following form

\begin{equation}
S[\hat\varphi,\hat{\varphi}^\dagger]
=
S_{\rm kin}
+
S_{\rm int}.
\end{equation}

The interaction term is, in general, combinatorially non-local in the
quantum geometric variables. At this stage, we also keep its dependence
on the scalar field variables general, with the corresponding couplings
encoded in the interaction kernel. We consider an interaction of the form

\begin{equation}
\begin{aligned}
S_{\rm int}
={}&
\int
\prod_{r=1}^{a}
\left[
d\chi_r\,d\eta_r\,dg_I^{(r)}
\right]
\prod_{s=1}^{b}
\left[
d\chi'_s\,d\eta'_s\,dg_I^{\prime(s)}
\right]
\\
&\times
\mathcal V_{a,b}
\left(
\{g_I^{(r)},\chi_r,\eta_r\};
\{g_I^{\prime(s)},\chi'_s,\eta'_s\}
\right)
\prod_{r=1}^{a}
\hat{\varphi}
\left(
g_I^{(r)},\chi_r,\eta_r
\right)
\prod_{s=1}^{b}
\hat{\varphi}^{\dagger}
\left(
g_I^{\prime(s)},\chi'_s,\eta'_s
\right)
+
\mathrm{h.c.},
\end{aligned}
\label{fundamental interaction}
\end{equation}

where $\mathcal V_{a,b}$ is the fundamental interaction kernel, encoding
both the combinatorial coupling of the geometric degrees of freedom and
the dependence of the interaction on the scalar-field variables.

The kinetic term is given by

\begin{equation}\label{S kin}
S_{\rm kin}
=
\int dg_I\, dg_I'\, d\chi\,d\chi'\,d\eta\ d\eta'\;
\hat{\varphi}^\dagger(g_I,\chi,\eta)
\,
\mathcal{K}(g_I, g_I', \chi,\chi', \eta, \eta')
\,
\hat{\varphi}(g_I',\chi',\eta')
\end{equation}

and it is assumed to respect the same symmetries of the Lagrangian density of a non-interacting, minimally coupled massless scalar field on an arbitrary background metric, that is it is invariant under the action of the Euclidean group via translation, reflection and rotations in field space. The reason is that this implies the same symmetries for the GFT Feynman amplitudes, which in turn will take the form of lattice path integrals for discrete gravity coupled to the same kind of discretized scalar fields \cite{TGFTscalarmatter}.


Therefore, in order to respect these symmetries, the kinetic operator takes the form

\begin{equation}
    \mathcal{K}(g_I,g_I',\chi,\chi',\eta,\eta') = \mathcal{K}\left(g_I,g_I',(\chi-\chi')^2+(\eta -\eta')^2\right),
\end{equation}

i.e. the derivative expansion of the kinetic operator can only involve powers of the Laplacian on the domain of the GFT field associated to such scalar matter. In principle, this expansion contains infinitely many derivative terms. Our use of coherent peaked states, in the following, will imply that we can truncate this expansion and retain terms up to second order only: 

\begin{equation}
S_{kin}
=
\int dg_I\, dg_I'\, d\chi\,d\eta \, \hat{\varphi}^\dagger(g_I,\chi,\eta)\left[ 
\mathcal{K}_0(g_I,g_I')+
\mathcal{K}_1(g_I,g_I')\,\left(\partial_\chi^2+ \partial_\eta^2\right)\right]\hat{\varphi}(g_I',\chi,\eta)
\end{equation}

where \(\mathcal{K}_i\) act only on geometric variables.

As anticipated, in the following, we restrict to the case \(G=SU(2)\). Then, by the Peter-Weyl theorem, the field operator can be expanded in terms of the matrix elements of the irreducible unitary representations of \(G\),

\begin{equation}
\hat{\varphi}(g_I,\chi,\eta)
=
\sum_J
\hat{\varphi}_J(\chi,\eta)
D^J(g_I),
\end{equation}

where $\pm J =  (j_I,\pm m_I, \kappa) $ collectively denotes spin and intertwiner labels and \(D^J(g_I)\) denotes the product of the matrix elements of the irreducible unitary representations of \(G\), while the multi-index \(J\) collectively labels the corresponding representation data.

The mode operators satisfy the commutation relations
\begin{equation}
\begin{aligned}
&[\hat{\varphi}_J(\chi,\eta),\hat{\varphi}^\dagger_{J'}(\chi',\eta')] = \delta_{JJ'} \delta(\chi - \chi')\delta(\eta - \eta')\\
&[\hat{\varphi}_J(\chi,\eta),\hat{\varphi}_{J'}(\chi',\eta')] = [\hat{\varphi}_J^\dagger(\chi,\eta),\hat{\varphi}_{J'}^\dagger(\chi',\eta')]= 0.
\end{aligned}
\end{equation}

The field operators are represented on a bosonic Fock space with Fock vacuum
\(|0\rangle\), satisfying
\begin{equation}
\hat{\varphi}_J(\chi,\eta)|0\rangle=0,
\end{equation}
for every \(J\), \(\chi\) and \(\eta\). Many-particle states are generated by repeated application of the creation operators
\(\hat{\varphi}_J^\dagger(\chi,\eta)\).

We consider \cite{Gielen:2020fgi} kinetic kernels that take a
diagonal form in the spin representation:

\begin{equation}
S_{\mathrm{kin}}
=
\sum_J
\int d\chi\,d\eta\,
\hat{\varphi}_J^\dagger(\chi,\eta)
\left[
K_0^{(J)}
+
K_1^{(J)}
\left(
\partial_\chi^2+\partial_\eta^2
\right)
\right]
\hat{\varphi}_J(\chi,\eta).
\end{equation}


It is also convenient to diagonalize with respect to the scalar field $\eta$, which will not be used as a relational clock when extracting the cosmological dynamics, by expressing the kernel and fields in terms of its momentum $\pi_{\eta}$ by Fourier transform:



\begin{equation}
S_{\mathrm{kin}}
=
\sum_J
\frac{1}{2\pi}
\int d\chi\,d\pi_\eta\,
\hat{\varphi}_J^\dagger(\chi,\pi_\eta)
\left[
K_0^{(J)}
+
K_1^{(J)}
\left(
\partial_\chi^2-\pi_\eta^2
\right)
\right]
\hat{\varphi}_J(\chi,\pi_\eta).
\end{equation}

For each mode $J$, we define

\begin{equation}\label{E^2 con pi}
E_J^2(\pi_\eta)
:=
-
\frac{K_0^{(J)}}{K_1^{(J)}} + \pi_\eta^2.
\end{equation}

Using this definition, the kinetic action can be written as

\begin{equation}
S_{\mathrm{kin}}
=
\sum_J
\frac{K_1^{(J)}}{2\pi}
\int d\chi\,d\pi_\eta\,
\hat{\varphi}_J^\dagger(\chi,\pi_\eta)
\left[
\partial_\chi^2
-
E_J^2(\pi_\eta)
\right]
\hat{\varphi}_J(\chi,\pi_\eta).
\end{equation}

In this second-quantised framework, extensive geometric observables correspond to one-body operators acting on the Fock space. For example, the number and volume operators are defined by

\begin{equation}\label{number operator}
\hat N
=
\sum_J
\int d\chi\,d\pi_\eta\,
\hat{\varphi}_J^\dagger(\chi,\pi_\eta)
\hat{\varphi}_J(\chi,\pi_\eta),
\end{equation}

\begin{equation}\label{volume operator}
\hat V
=
\sum_J
V_J
\int d\chi\,d\pi_\eta\,
\hat{\varphi}_J^\dagger(\chi,\pi_\eta)
\hat{\varphi}_J(\chi,\pi_\eta),
\end{equation}

where $V_J$ denotes the volume eigenvalue associated with a single GFT quantum labelled by $J$.



The effective cosmological dynamics is then obtained in a few steps. 
First, we take a mean field approximation of the fundamental quantum GFT dynamics; in turn, this corresponds to restricting attention to a suitable family of simple coherent condensate states; notice that this corresponds to taking a very coarse, approximate continuum limit of the full theory. 
Next, since our goal is to describe homogeneous and isotropic cosmological spacetimes, we further
restrict our attention to the isotropic sector of such states and of the corresponding mean field dynamics. 
Then, by a further restriction of such coherent condensate states to those with good peaking properties on the values of the scalar degree of freedom we want to use as a relational clock, we can turn the general mean field equations into evolution equations with respect to these peak values of the matter clock, i.e. our physical time variable. 
Finally, we can derive equations of motion for the relevant observables, in particular the \lq total volume of the universe\rq, taken in their expectation values in the same (isotropic, peaked) condensate states; the resulting dynamics has then the interpretation of the cosmological dynamics we want to compare with classical gravitational physics. 
\\
\\
\\
\vspace{0.5cm}

{\bf Coherent peaked states and mean field dynamics}

In the isotropic sector,
the geometric data of each GFT quantum describe equilateral tetrahedra,
so that the four spin labels coincide,
\[
J=(j,j,j,j,\kappa).
\]

For each spin $j$, we further fix the intertwiner label $\kappa$
by selecting the state corresponding to the maximal volume eigenvalue
compatible with the four equal face areas. With this choice, the
geometric multi-index reduces to the single geometric (spin) label $j$, as we expect from isotropic (and not space-dependent) geometries.

The corresponding coherent condensate states
are

\begin{equation}
|\sigma\rangle
=
\mathcal{N}(\sigma)
\exp\left[
\sum_j \frac{1}{2\pi}
\int d\chi\, d\pi_\eta \,
\sigma_j(\chi,\pi_\eta)
\hat{\varphi}_j^\dagger(\chi,\pi_\eta)
\right]
|0\rangle ,
\end{equation}

where \(|0\rangle\) denotes the Fock vacuum and
\(\mathcal{N}(\sigma)\) is a normalisation factor. These states satisfy

\begin{equation}
\hat{\varphi}_j(\chi,\pi_\eta)
|\sigma\rangle
=
\sigma_j(\chi,\pi_\eta)
|\sigma\rangle ,
\end{equation}

so that the complex function
\(\sigma_j(\chi,\pi_\eta)\) plays the role of the condensate mean field.

Then, we restrict the coherent states to be sharply localised both in
\(\chi\) and in the momentum \(\pi_\eta\). The localisation around a value \(\chi=\chi_0\) singles out
\(\chi\) as a relational clock, identifying \(\chi_0\) with a relational
instant so that one gets \lq\lq time derivatives\rq\rq{} with respect to it from the $\chi$-derivatives in the GFT action. The localisation around a fixed value
\(\pi_\eta=\bar{\pi}_\eta\), instead, selects a semiclassical matter sector
which, in the non-interacting theory, has been shown to reproduce the FLRW dynamics \cite{Marchetti:2021gcv}. The condensate wave function is
therefore assumed to take the form

\begin{equation}
\sigma_j(\chi,\pi_\eta)
=
\delta_\epsilon(\chi-\chi_0)
\,
\delta_\Delta(\pi_\eta-\bar{\pi}_\eta)
\,
\tilde{\sigma}_j(\chi,\pi_\eta),
\end{equation}

where \(\delta_\epsilon\) and \(\delta_\Delta\) are localisation functions
centred around \(\chi_0\) and \(\bar{\pi}_\eta\), with dispersion $\epsilon$ and $\Delta$, respectively, while
\(\tilde{\sigma}_j(\chi,\pi_\eta)\) denotes the $reduced$ $condensate$ $wave$ $function$, assumed not to modify the peaking behaviour imposed by the
localisation profiles.
A simple choice for the localisation functions is given by Gaussian profiles. 


The restriction to coherent peaked states specifies the sector in
which the effective cosmological dynamics will be studied, and it has also the effect of allowing the approximation of a generic higher-derivative kinetic kernel with the second-order form introduced above.

In the sharp localisation limit, the condensate wave function is effectively characterised by the reduced mean field

\begin{equation}
\tilde{\sigma}_j(\chi,\pi_\eta)
\longrightarrow
\tilde{\sigma}_j(\chi_0,\bar{\pi}_\eta)
\end{equation}

and the contribution of the additional scalar field enters the effective dynamics through $E_j^2$ introduced in Eq.~\eqref{E^2 con pi}.

In the following, we restrict our attention to the sector
$E_j^2>0$, since this has been shown to correspond to exponentially growing solutions
in relational time, i.e. the relevant ones for cosmological dynamics.
The complementary sector $E_j^2<0$ corresponds instead to oscillatory
solutions and will not be considered.\\
The expectation values of the number and volume operators defined in Eqs.~\eqref{number operator} and \eqref{volume operator} on peaked coherent condensate states provide the corresponding relational observables \cite{Oriti:2016qtz}, since they introduce a dependence on clock-time values of the a priori time-independent operators:

\begin{equation}\label{N operator sigma}
N(\chi_0, \bar \pi_{\eta})
:=
\langle\sigma|
\hat N
|\sigma\rangle
\simeq
\sum_j
\left|
\tilde{\sigma}_j(\chi_0,\bar{\pi}_\eta)
\right|^2,
\end{equation}

\begin{equation}\label{V operator sigma}
V(\chi_0, \bar\pi_\eta)
:=
\langle\sigma|
\hat V
|\sigma\rangle
\simeq
\sum_j
V_j
\left|
\tilde{\sigma}_j(\chi_0,\bar{\pi}_\eta)
\right|^2,
\end{equation}

where $V_j$ denotes the volume eigenvalue associated with a single isotropic GFT quantum labelled by the spin $j$.

Clearly, the dynamical equations for the reduced GFT mean field will induce dynamical equations for these relational observables, since the mean field enters in their definition, as specified above.\\
The effective mean field dynamics is obtained by evaluating the quantum equations of motion on the condensate coherent state and neglecting quantum fluctuations around the mean field, that is considering only the expectation value of the quantum equations of motion in the chosen condensate state. This leads to

\begin{equation}
\left\langle \sigma \left|
\frac{\delta S}
{\delta \hat{\varphi}_j^\dagger}
\right|\sigma\right\rangle
=
0,
\end{equation}

which yields the effective mean-field equation

\begin{equation}\label{effectiveEOM}
\frac{\delta S[\tilde{\sigma},\tilde{\sigma}^*]}
{\delta \tilde{\sigma}_j^*}
=
0.
\end{equation}


\vspace{0.5cm}

{\bf Mean field equations}

The mean-field equation~\eqref{effectiveEOM} is obtained by projecting the
fundamental GFT dynamics onto the isotropic coherent peaked states introduced
above. Under this reduction, the kinetic term gives rise to the linear
operator
\begin{equation}
L_j
=
\partial_{\chi_0}^2-E_j^2,
\end{equation}

while the fundamental interaction term~\eqref{fundamental interaction} gives,
instead, a nonlinear contribution determined by the projection of the
interaction kernel $\mathcal V_{a,b}$ onto the same reduced condensate
sector. Denoting the resulting effective interaction potential for each
isotropic mode by $\mathcal V_j$, the mean-field equations take the form
\begin{equation}
L_j[\tilde{\sigma}_j]
+
U_j[\tilde{\sigma}_j,\tilde{\sigma}_j^*]
=
0,
\qquad
U_j
=
\frac{\partial\mathcal V_j}
{\partial\tilde{\sigma}_j^*}.
\end{equation}

The effective interaction obtained from the fundamental interaction introduced
above has the same general structure as the class of interactions considered
in Ref.~\cite{Marchetti:2025jze}. In contrast with the non-Hermitian case
studied there, however, in the present work we focus on Hermitian interactions.
The resulting effective interaction potential for each isotropic mode is
therefore

\begin{equation}
\mathcal V_j[\tilde{\sigma}_j,\tilde{\sigma}_j^*]
=
\gamma_j
\tilde{\sigma}_j^a
(\tilde{\sigma}_j^*)^b
+
\gamma_j^*
(\tilde{\sigma}_j^*)^a
\tilde{\sigma}_j^b.
\label{effective interaction potential}
\end{equation}
Here, $\gamma_j$ is the effective complex coupling resulting from the
projection of the fundamental interaction kernel onto the reduced isotropic
condensate sector.

Introducing the parametrisation

\begin{equation}
    a = \frac{l+m+1}{2}; \hspace{1cm} b = \frac{l-m+1}{2}; \hspace{1cm} \gamma_j=-\frac{\lambda_j}{b}e^{i \varphi_j}; \hspace{1cm} c_j = \frac{a}{b}e^{-2i\varphi_j},
\end{equation}
where $\varphi_j, \lambda_j \in \mathbb{R}$ and using the polar decomposition in terms of standard hydrodynamic variables (fluid density and phase) $\tilde{\sigma}_j = \rho_j e^{i\theta_j}$, we obtain 

\begin{equation}
    U_j = -\lambda_j\rho_j^l e^{i\varphi_j}[e^{i(m+1)\theta_j} + c_j e^{i(1-m)\theta_j}].
\end{equation}

Separating the equation into its real and imaginary parts then yields

\begin{equation}\label{eq motion rho'' e theta''}
\left\{
\begin{array}{l}
\rho_j''- \left[ (\theta_j')^2+E_j^2 \right] \rho_j
-\dfrac{2\lambda_j(l+1)}{l-m+1}\rho_j^l
\cos(\varphi_j+m\theta_j)=0,
\\[10pt]
\rho_j\theta_j'' +2\rho_j'\theta_j'
+\dfrac{2\lambda_j m}{l-m+1}\rho_j^l
\sin(\varphi_j+m\theta_j)=0.
\end{array}
\right\} ,
\end{equation}

where a prime denotes differentiation with respect to the relational time $\chi_0$.\\
Note that requiring \(a\in\mathbb{N}_0\) and \(b\in\mathbb{N}^{+}\) implies
\[
l\in\mathbb{N}_0,\qquad
m\in\mathbb{Z},\qquad
-l-1\le m\le l-1,\qquad
l+m\equiv1\pmod 2,
\]
where the last condition simply states that \(l\) and \(m\) must have opposite parity.\\
Since the action is invariant under translations in the relational time $\chi_0$, the associated conserved Noether charge, which we refer to as the $GFT$ $energy$, is given by

\begin{equation}\label{GFT energy}
\mathcal{E}_j =
\frac{1}{2} (\rho_j')^2
+\frac{1}{2} \rho_j^2 (\theta_j')^2
-\frac{1}{2} E_j^2 \rho_j^2
-\frac{2\lambda_j}{l-m+1}\rho_j^{l+1}\cos(\varphi_j+m\theta_j).
\end{equation}

Notice that the condensate modes decouple, under our assumptions and choice of microscopic model, in that we have an independent equation for each mode. On the other hand, different condensate modes contribute to the expression for the cosmological observables, in particular the universe volume, and thus to their dynamics.\\
Having extracted the mean field dynamics of the system, from which we can easily derive the dynamical equations for the universe volume observable, we can start our analysis of them.


\section{Asymptotic dynamics}\label{sec: Asymtotic dynamics}

We now investigate the asymptotic dynamics of the condensate modes described
by Eq.~\eqref{eq motion rho'' e theta''}. In general, as remarked above, the condensate contains
contributions from several spin modes, with the total volume given by
Eq.~\eqref{V operator sigma}. 

For each mode, the dynamics of the amplitude
$\rho_j$ and of the phase $\theta_j$ are coupled through
Eq.~\eqref{eq motion rho'' e theta''}. Consequently, although the volume
depends explicitly only on the condensate amplitudes, the phase dynamics also
contributes to its evolution through its coupling to $\rho_j$.

Since the equations of motion for different spin modes are decoupled, their
asymptotic dynamics can be studied independently. To study the asymptotic
($\rho_j \gg 1$) stability properties of each mode, relevant for the large-volume, late-time cosmological evolution, we introduce the
variables
\begin{equation*}
x_j=\theta_j,\qquad
y_j=\frac{\rho_j'}{\rho_j^3},\qquad
z_j=\theta_j'.
\end{equation*}

Substituting these variables into \eqref{eq motion rho'' e theta''}, we obtain the dynamical system

\begin{equation}\label{x', y', z'}
\left\{
\begin{array}{l}
x_j' = z_j, \\[6pt]
y_j' = -3\rho_j^2 y_j^2 + \dfrac{z_j^2 + E_j^2}{\rho_j^2}
+\dfrac{2\lambda_j(l+1)}{l-m+1}\rho_j^{l-3}
\cos(\varphi_j + mx_j), \\[8pt]
z_j' = -2\dfrac{\rho_j'}{\rho_j}z_j
-\dfrac{2\lambda_j m}{l-m+1}\rho_j^{\,l-1}
\sin(\varphi_j + mx_j).
\end{array}
\right.
\end{equation}

The corresponding fixed points are given by

\begin{equation}\label{fixed_points}
\left\{
\begin{array}{l}
\bar{x}_{j,n}=\dfrac{-\varphi_j+n\pi}{m},
\qquad n\in\mathbb{Z}, \\[10pt]

\bar{y}_{j,n}^2=
\dfrac{1}{3}
\left[
\dfrac{E_j^2}{\rho_j^4}
+\dfrac{2\lambda_j(l+1)}{l-m+1}
(-1)^n\rho_j^{\,l-5}
\right]
\xrightarrow{\rho_j\to+\infty}
\dfrac{2\lambda_j(l+1)}{3(l-m+1)}
(-1)^n\rho_j^{\,l-5}, \\[12pt]

\bar{z}_j=0.
\end{array}
\right.
\end{equation}

It should be noted that this excludes the special case $m=0$. However, this does not represent a limitation, as in this special case, the present model exactly reduces to the one previously studied in \cite{Pang:2025jtk}, to which we refer for the corresponding analysis and results.
In the following, we denote the corresponding fixed point simply by
$(\bar{x}_j,\bar{y}_j,\bar{z}_j)$, omitting the index $n$.

In order for $\bar{y}_{j}^2$ to be positive, we select the branch of fixed points
for which
\[
(-1)^n\operatorname{sgn}(\lambda_j)>0.
\]

Furthermore, requiring the fixed point to remain finite in the large-volume limit
implies $l\leq5$, while a non-vanishing asymptotic value is obtained only for
$l=5$. Therefore,
\begin{equation}\label{y^bar^2}
    \bar{y}_j^{\,2}
    =\frac{4}{6-m}\,|\lambda_j|.
\end{equation}

As shown in the following, this asymptotic fixed point corresponds to an emergent de Sitter phase.

Following \cite{Marchetti:2025jze}, we now investigate the linear stability of the selected fixed point for each mode by introducing the perturbations
\begin{equation*}
x_j=\bar{x}_j+\xi_j,\qquad
y_j=\bar{y}_j+\iota_j,\qquad
z_j=\bar{z}_j+\zeta_j.
\end{equation*}
To first order, the perturbations satisfy the linearized system
\begin{equation}
\begin{pmatrix}
\xi_j' \\
\iota_j' \\
\zeta_j'
\end{pmatrix}
=
\mathcal{J}_j
\begin{pmatrix}
\xi_j \\
\iota_j \\
\zeta_j
\end{pmatrix},
\end{equation}
where \(\mathcal{J}_j\) denotes the Jacobian matrix of the dynamical system
evaluated at the fixed point,
\begin{equation}
\mathcal{J}_j
=
\begin{pmatrix}
0 & 0 & 1 \\[6pt]
0 & -6\rho_j^2\bar{y}_j & 0 \\[8pt]
-\dfrac{m^2}{2}\rho_j^4\bar{y}_j^2 & 0 & -2\rho_j^2\bar{y}_j
\end{pmatrix}.
\end{equation}

Integrating the equation for $\iota_j$ yields
\begin{equation}\label{iota linear solution}
    \iota_j=\frac{C_j}{\rho_j^6},
\end{equation}
where $C_j$ is an integration constant.\\
Using the definition $\tau_j=\rho_j^{-2}$, the equation for $\xi_j$ can be
rewritten as
\begin{equation}
    \tau_j^2 \frac{d^2\xi_j(\tau_j)}{d\tau_j^2}
    - \tau_j \frac{d\xi_j(\tau_j)}{d\tau_j}
    + \frac{m^2}{8}\xi_j(\tau_j)=0 \quad . 
\end{equation}
Looking for solutions of the form
$\xi_j(\tau_j)=\tau_j^r$ yields the characteristic exponents
\begin{equation}
    r=1\pm\sqrt{1-\frac{m^2}{8}}
    \equiv 1\pm\mu.
\end{equation}

Accordingly, the solution takes one of the following forms:
\begin{equation}\label{xi(rho)}
\xi_j(\rho_j)=
\frac{1}{\rho_j^{2}}
\begin{cases}
A_j\rho_j^{-2\mu}+B_j\rho_j^{2\mu},
&
1-\dfrac{m^2}{8}>0,
\\[6pt]

A_j+B_j\log\rho_j,
&
1-\dfrac{m^2}{8}=0,
\\[6pt]

A_j\cos(\phi_j)+B_j\sin(\phi_j),
&
1-\dfrac{m^2}{8}<0,
\end{cases}
\end{equation}
where
\[
\phi_j(\rho_j)\equiv\beta\log\rho_j,
\qquad
\beta\equiv2\sqrt{-1+\frac{m^2}{8}}.
\]

The corresponding solution for $\zeta_j$ is then obtained from the linearized
equations:
\begin{equation}\label{zeta(rho)}
\zeta_j(\rho_j)=
-2\bar{y}_j
\begin{cases}
A_j(1+\mu)\rho_j^{-2\mu}
+
B_j(1-\mu)\rho_j^{2\mu},
&
1-\dfrac{m^2}{8}>0,
\\[6pt]

A_j-\dfrac{B_j}{2}
+
B_j\log\rho_j,
&
1-\dfrac{m^2}{8}=0,
\\[6pt]

c_{A,j}\cos(\phi_j)
+
c_{B,j}\sin(\phi_j),
&
1-\dfrac{m^2}{8}<0,
\end{cases}
\end{equation}
where
\[
c_{A,j}\equiv A_j-\frac{\beta B_j}{2},
\qquad
c_{B,j}\equiv B_j+\frac{\beta A_j}{2}.
\]

Although the interaction parameter $m$ is restricted to integer values in the present model, the above analysis has been carried out for arbitrary real $m$, yielding a complete classification of the asymptotic solutions. The physically relevant cases are then obtained by restricting to the allowed integer values of $m$. In particular, the case $1-m^2/8=0$ is never realized. Throughout the remainder of this work, $m$ will be treated as a real parameter in order to study each regime in full generality, while keeping in mind that only the integer values are relevant within the present model.\\

In the oscillatory regime, corresponding to $1 - m^2/8 < 0$, the $\xi_j$-component of the perturbation decays as $\rho_j^{-2}$, whereas the $\zeta_j$-component remains bounded and oscillates with increasing frequency. This asymptotic behaviour is identical to that found in \cite{Marchetti:2025jze}. Following the analysis presented therein, such fixed points can therefore be interpreted as \emph{local weak forward attractors}, in the sense that the rapidly oscillating perturbations become asymptotically indistinguishable from zero under a coarse-grained (weak) description. 

It is important to stress that the above conclusions about the asymptotic behaviour are valid {\it for all condensate modes $j$}, independently of each other.

\vspace{0.5cm}

{\bf Single-mode asymptotic regime}

So far we have considered the independent contribution of different condensate modes to the
total volume in the asymptotic regime. For an expanding mode approaching the
fixed point, the definition of $y_j$ gives
\begin{equation}
    \frac{\rho_j'}{\rho_j^3}
    \simeq \bar y_j ,
\end{equation}
with $\bar y_j>0$. The corresponding leading order solution is
\begin{equation}\label{leading rho solution}
    \rho_{j,0}^2(\chi_0)
    =
    \frac{1}{
    2\bar y_j
    \left(
    \chi_{0,j}^{(\infty)}-\chi_0
    \right)
    },
\end{equation}
where $\chi_{0,j}^{(\infty)}$ denotes the relational time at which the
leading order solution of the $j$-th mode diverges.

If different modes are characterized by distinct values of
$\chi_{0,j}^{(\infty)}$, the mode reaching its asymptotic divergence first
becomes dominant, namely the mode with the smallest
$\chi_{0,j}^{(\infty)}$. Indeed, denoting this mode by $j=1$, as
$\chi_0\rightarrow\chi_{0,1}^{(\infty)}$ one has
$\rho_1^2\rightarrow\infty$, while the amplitudes of the other modes remain
finite. Consequently,
\begin{equation}\label{LABEL_MODE_DOMINANCE}
    \frac{V_j\rho_j^2}{V_1\rho_1^2}
    \longrightarrow 0,
    \qquad j\neq 1,
\end{equation}

and the expectation values of the number and volume operators in
Eqs.~\eqref{N operator sigma} and \eqref{V operator sigma}, respectively,
asymptotically reduce to
\begin{equation}
    N(\chi_0)\simeq \rho_1^2(\chi_0),
\end{equation}

\begin{equation}
    V(\chi_0)\simeq V_1\rho_1^2(\chi_0).
\end{equation}

Thus, sufficiently close to the large-volume asymptotic regime, the
contribution of the dominant mode overwhelms those of the remaining
condensate modes, and the cosmological dynamics becomes effectively
single-mode. Then, the asymptotic behaviour of the relevant
cosmological observables is entirely determined by the dominant mode.
The single-mode approximation is therefore not imposed a priori, but
emerges dynamically as a consistent description of the asymptotic regime.\\
In the following, we shall exploit this result to study the corresponding
late-time cosmological dynamics, omitting the spin label of the dominant
mode for notational simplicity.

\vspace{0.5cm}

{\bf Emergent cosmological parameters}

In a homogeneous and isotropic universe, the matter content can be described by a perfect fluid satisfying the equation of state
\begin{equation}
    p = w\varrho,
\end{equation}
where $p$ and $\varrho$ denote the pressure and energy density, respectively. Different values of the equation-of-state parameter $w$ correspond to different cosmological evolutions.

Following the construction of \cite{Oriti:2021rvm, PithisSakellariadou}, the effective equation of state parameter and the Hubble rate, expressed in terms of the relational clock, are given by

\begin{equation}\label{espressione w}
    w = -1-\frac{y'}{\rho^2y^2},
\end{equation}

and

\begin{equation}
    H^2
    =
    \left(
    \frac{1}{3}\frac{V'}{V^2}\pi_\chi
    \right)^2
    =
    \frac{4}{9}\frac{\pi_\chi^2}{V_1^2}y^2,
\end{equation}

where $\pi_\chi$ denotes the momentum associated with the
clock scalar field $\chi$, which is constant in the cosmic time gauge.

Here we have used the crucial result that, at very late times, thus large condensate densities and universe volumes, eventually a single mode dominates the evolution of the system. Therefore, {\it asymptotically}, the expression for the universe volume and the effective equation of state are fully captured by the contribution of this single dominant mode. This allows for the simplified expression above. When considering the more general issue of the {\it approach} to the late-time asymptotic regime, thus the issue of a dynamical dark energy, in general one should account properly for the contribution of different condensate modes.\\
Evaluating these quantities at the fixed point, where $H^2=\Lambda/3$, yields
\begin{equation}
    \bar{w}=-1,
\end{equation}
corresponding to an asymptotic de Sitter phase, and an effective cosmological constant
\begin{equation}
    \bar{\Lambda}
    =
    \frac{16}{3}
    \frac{\pi_\chi^2}{V_1^2}
    \frac{\left|\lambda\right|}{6-m}.
\end{equation}

Interestingly, unlike the model in \cite{Marchetti:2025jze}, here the effective cosmological constant explicitly depends on the interaction parameter $m$ (notice that $-6\leq m\leq4$ for $l=5$). This difference can be traced back to the different interactions considered in the two models. In \cite{Marchetti:2025jze}, for generic non-Hermitian interactions, the parameter $m$ enters the equations of motion through the phase dependence of the trigonometric interaction terms, but does not affect their strength.
In the present model, requiring the interaction to be Hermitian for generic admissible values of $m$ makes the strength of the interaction terms in the equations of motion explicitly dependent on $m$, as can be seen from Eq.~\eqref{eq motion rho'' e theta''}. This $m$-dependence persists at the asymptotic fixed point, yielding $\bar y^2=4|\lambda|/(6-m)$, and is therefore inherited by the effective cosmological constant.


\vspace{0.5cm}

{\bf Stability of the fixed points}

The role of the interaction parameter $m$ can be understood by analysing the phase equation \eqref{eq motion rho'' e theta''},
\begin{equation}
\theta'' +2\frac{\rho'}{\rho}\theta'
+\frac{2\lambda m}{l-m+1}\rho^{\,l-1}\sin(\varphi+m\theta)=0 .
\end{equation}
The evolution of the phase is governed by the competition between the nonlinear interaction term and the dissipative contribution $2(\rho'/\rho)\theta'$. For $\rho'>0$, corresponding to an expanding universe, the latter acts as a friction term that continuously dissipates the kinetic energy of the phase. Consequently, as the universe expands, $\theta$ is no longer able to overcome neighbouring potential barriers and is driven towards the nearest stationary configuration.

The stationary points are determined by
\[
\sin(\varphi+m\bar{\theta}_n)=0,
\]
namely
\begin{equation}
\bar\theta_n=\frac{-\varphi+n\pi}{m},
\end{equation}

which is precisely the fixed point value previously obtained from the first equation of the dynamical system \eqref{fixed_points}, here expressed in terms of the phase variable $\theta$.\\
To investigate their stability, we linearize the phase equation around a stationary point by writing
$\theta=\bar\theta_n+\delta$. At first order, one obtains
\begin{equation}
\delta''+2\frac{\rho'}{\rho}\delta'
+\frac{2\lambda(-1)^n m^2}{l-m+1}\rho^{\,l-1}\delta=0 .
\end{equation}
The effective restoring coefficient is therefore proportional to
$\lambda(-1)^n m^2/(l-m+1)$. After selecting the physical branch satisfying $\bar y^2>0$, this coefficient is always positive, implying that the corresponding stationary point is dynamically stable.

The sign of $m$ does not affect the stability of the phase configurations,
but its value determines both the location of the stationary points:
$\bar{\theta}_n=(-\varphi+n\pi)/m$, and the way in which they are approached
asymptotically. In the non-oscillatory regime, Eq.~\eqref{xi(rho)} shows that
$\xi\to0$, so that the phase approaches its stationary value, while
Eq.~\eqref{zeta(rho)} shows that the corresponding perturbation $\zeta$ in
the phase velocity can grow asymptotically. In the oscillatory regime,
instead, Eq.~\eqref{xi(rho)} shows that $\xi$ decays as $\rho^{-2}$, whereas
Eq.~\eqref{zeta(rho)} shows that $\zeta$ remains bounded and oscillatory,
reproducing the weak-attractor behaviour discussed in
Ref.~\cite{Marchetti:2025jze}. This differs from the model studied there,
where the sign of $m$ determines the stability of the stationary points.\\
Therefore, the emergent de Sitter phase is associated with an asymptotically stable fixed point of the condensate dynamics for all admissible values of the interaction parameter $m$. One physical implication is that our model can describe a viable late-time cosmology, of interest for dark energy modelling, but it cannot produce an unstable inflationary phase (i.e. one with slow roll and graceful exit) in the early universe, contrary to the non-hermitian interactions considered in \cite{Marchetti:2025jze}.


\section{Dynamical Dark Energy} \label{sec:dynamical-dark-energy}

In the previous section, we characterized the strict asymptotic regime of the
model. We showed that, under the conditions discussed above, one condensate
mode eventually dominates the total volume, so that the late-time cosmological
dynamics becomes effectively single-mode and approaches a de Sitter phase,
with $w\to-1$.

We now turn to the approach to this asymptotic regime. At finite, although
large, volume, contributions that become negligible in the strict asymptotic
limit can still affect the cosmological evolution and, in particular, the
effective equation of state. Two distinct effects are relevant in this
respect. First, within each condensate mode, the phase dynamics affects the
evolution of the corresponding amplitude through the coupled equations of
motion \eqref{eq motion rho'' e theta''}. Second, before the asymptotic
single-mode regime is reached, subdominant condensate modes can still
contribute to the total volume in Eq.~\eqref{V operator sigma}.

In particular, although the contribution of the subdominant modes vanishes
relative to that of the dominant mode in the strict asymptotic limit, as shown
in Eq.~\eqref{LABEL_MODE_DOMINANCE}, it can still influence the cosmological
dynamics during the approach to this limit. We capture the leading multimode
correction by retaining, in addition to the asymptotically dominant mode, a
second condensate mode. The dynamical dark-energy behaviour can therefore
receive contributions from both the phase dynamics and the presence of a
subdominant condensate mode.

We study these two effects progressively. We first restrict again to the
single-mode sector and investigate how the phase dynamics modifies the
approach to the asymptotic de Sitter regime. In this case, using the
single-mode equation of state parameter defined in Eq.~\eqref{espressione w},
we write its asymptotic behaviour as
\begin{equation}\label{w(rho)}
w(\rho)=-1+\delta w(\rho),
\end{equation}
where
\begin{equation}\label{deltaw(rho)}
\delta w(\rho)
=
-\frac{y'}{\rho^2y^2}.
\end{equation}

As shown in the previous section, the asymptotic behaviour of the phase
perturbations depends qualitatively on the interaction parameter $m$.
We therefore study separately the oscillatory and non-oscillatory regimes,
and determine in each case the corresponding phase-induced corrections to
the equation of state. In the next section, we will go beyond the single-mode
description by including a second condensate mode and studying its contribution
to the dynamical dark-energy evolution, together with its interplay with the
phase dynamics.


\vspace{0.5cm}

{\bf Oscillatory regime: $1-m^2/8<0$}
\label{oscillatory regime in Sec 4}

For $l=5$, the oscillatory regime occurs only for the allowed values
$m=-6,\pm4$.\\ As shown in the previous section, in this regime the phase
perturbations $\xi$ and $\zeta$ exhibit an oscillatory asymptotic behaviour,
given by Eqs.~\eqref{xi(rho)} and \eqref{zeta(rho)}.

In order to determine the leading correction to the equation of state, the
linear analysis is not sufficient, and the evolution equation for $y$ must be
expanded up to second order in the perturbations. The resulting asymptotic
solution takes the form
\begin{equation}\label{y oscillatory asymptotic}
y
=
\bar y
\left(
1+\frac{g(\phi)}{\rho^4}
+\frac{C}{\rho^6}
\right),
\end{equation}
where $\phi=\beta\log\rho$ and $C$ is an integration constant, and $g(\phi)$ is a periodic function given by
\begin{align}
g(\phi)
=\;&
\frac{E^2}{2\bar y^2}
-\frac{k_A+k_B}{4}
+
\frac{-k_A+k_B+\beta k_{AB}}
{4(1+\beta^2)}
\cos 2\phi
+
\frac{-\beta k_A+\beta k_B-k_{AB}}
{4(1+\beta^2)}
\sin 2\phi ,
\label{g oscillatory}
\end{align}
with
\begin{equation}
k_A
=
\frac{3}{2}m^2A^2-4c_A^2,
\qquad
k_B
=
\frac{3}{2}m^2B^2-4c_B^2,
\qquad
k_{AB}
=
3m^2AB-8c_Ac_B.
\end{equation}
The details of the perturbative calculation leading to these expressions are
reported in Appendix~\ref{app:oscillatory-regime}.

Substituting Eq.~\eqref{y oscillatory asymptotic} into
Eq.~\eqref{deltaw(rho)}, one obtains

\begin{equation}\label{deltaw esplicita}
\delta w(\rho)
=
\frac{1}{\rho^4}
\left[
\frac{2E^2}{\bar y^2}
+h(\phi)
+\frac{6C}{\rho^2}
\right],
\end{equation}

where the oscillatory contribution can be written as

\begin{equation}
h(\phi)
=
\delta w_0
+\delta w_1\cos2\phi
+\delta w_2\sin2\phi,
\end{equation}
with
\begin{equation}
\delta w_0
\equiv
-(k_A+k_B),
\end{equation}
\begin{equation}
\delta w_1
\equiv
-\frac{
(2-\beta^2)(k_A-k_B)-3\beta k_{AB}
}{
2(1+\beta^2)
},
\end{equation}
\begin{equation}
\delta w_2
\equiv
-\frac{
3\beta(k_A-k_B)+(2-\beta^2)k_{AB}
}{
2(1+\beta^2)
}.
\end{equation}

One can show that the maximum of the function $h(\phi)$ is zero:
\begin{equation}
h_{\max}
=
\delta w_0
+
\sqrt{\delta w_1^2+\delta w_2^2}
=
0,
\end{equation}
for any real values of $A$ and $B$, and therefore
\begin{equation}
h(\phi)\leq0.
\end{equation}

Neglecting the subleading contribution proportional to $C$, a phantom crossing can occur only if the negative oscillating contribution $h(\phi)$ overcomes the positive constant term $2E^2/\bar{y}^2$. Thus, the occurrence of the phantom crossing depends on the specific asymptotic trajectory of the condensate.

In principle, this condition could be further investigated by comparing the minimum of $h(\phi)$ with the constant contribution $2E^2/\bar{y}^2$. However, the integration constants $A$ and $B$ characterize the perturbative asymptotic expansion rather than the full cosmological evolution. At present, we do not know how to relate these asymptotic parameters to the global initial conditions of the condensate dynamics, preventing a general criterion for the existence of a phantom crossing.\\

An interesting implication follows from the decomposition in
Eq.~\eqref{E^2 con pi}, since the constant positive contribution entering
Eq.~\eqref{deltaw esplicita} is proportional to \(E^2\), an
increase of \(\bar\pi_\eta\) suppresses the possibility that the oscillatory
term \(h(\phi)\) overcomes it, thereby making the phantom regime
progressively harder to realize. Consequently, if one requires the
cosmological evolution to be compatible with recent observational
indications of dynamical dark energy involving a crossing of the phantom
divide \cite{DESI:2025fii,Ozulker2025PhantomCrossing}, one is no longer free to choose the conserved momentum \(\bar\pi_\eta\) arbitrarily large. Rather, the condition for the existence of a phantom crossing implies, at least in principle, the existence of an upper bound on the allowed values of \(\bar\pi_\eta\), which is a proxy for the total matter energy content of the universe, in our model in which this is only given by such matter field. Although such a bound cannot yet be determined
quantitatively without relating the asymptotic parameters \(A\) and \(B\)
to the global initial conditions of the condensate dynamics, this result
suggests that observational constraints on the dark-energy equation of
state may be translated, in our model, into constraints on the (conserved) momentum and thus energy density of matter in the universe.


\vspace{0.5cm}

{\bf Non-oscillatory regime: $1-m^2/8\geq0$}

For $l=5$, among the non-zero admissible values of $m$, this regime is
realized for $m=\pm2$.\\ As shown in the previous section, the asymptotic
phase perturbations are characterized by
\begin{equation}
\mu=\sqrt{1-\frac{m^2}{8}},
\end{equation}
with $\mu\in[0,1)$ when $m$ is treated as a real parameter.

To determine the correction to the equation of state, we
expand around the asymptotic fixed point. The details of the
calculation are reported in Appendix~\ref{app:non-oscillatory-regime}.
At leading order, one finds
\begin{equation}
y(\xi)
=
\bar y\left(1+y^*m^2\xi^2\right),
\qquad
y^*
=
-\frac{1}{2(1+\mu)}.
\end{equation}

The resulting correction to the equation of state parameter is
\begin{equation}
\delta w(\rho)
=
\frac{\delta w_0}{\rho^4}
\begin{cases}
\rho^{4\mu},
&
0<\mu<1,
\\[8pt]
\log^2\rho,
&
\mu=0,
\end{cases}
\label{deltaw_nonosc}
\end{equation}
where
\begin{equation}
\delta w_0
=
-16(1-\mu)^2B^2.
\end{equation}

Since $\delta w_0$ is always negative, the correction remains negative
throughout the non-oscillatory regime,
\begin{equation}
\delta w(\rho)<0.
\end{equation}
At the same time, for $\mu\in[0,1)$ the correction vanishes asymptotically,
so that $w\to-1$. The de Sitter fixed point is therefore approached from
the phantom side.

\vspace{0.5cm}

{\bf Phantom crossing}

We now investigate more specifically the conditions for a phantom crossing,
which occurs when the equation of state parameter crosses the value

\begin{equation}
w=-1,
\end{equation}

which corresponds to

\begin{equation}
y'
=
\frac{1}{\rho^4}
\left(
\rho\rho''
-
3(\rho')^2
\right)
=
0.
\end{equation}

Using the first equation in \eqref{eq motion rho'' e theta''} together with the positive solution for $\rho'$ obtained from \eqref{GFT energy}, corresponding to the expanding cosmological solution, we obtain

\begin{equation}\label{analytic expansion}
w+1
=
\delta w
=
\frac{
2\rho^2\left(E^2-2(\theta')^2\right)
+
6\mathcal E
}
{\rho^6y^2},
\end{equation}

which immediately yields the crossing condition

\begin{equation}\label{rho_c vs theta_c}
\rho_c^2
=
\frac{3\mathcal E}
{2(\theta_c')^2-E^2},
\end{equation}

where $\chi_c$ is the relational time at which the phantom crossing occurs, with $\rho_c\equiv\rho(\chi_c)$ and $\theta_c'\equiv\theta'(\chi_c)$.\\
Equation~\eqref{rho_c vs theta_c} provides an exact relation between the condensate amplitude and the phase velocity at the phantom crossing, without relying on any asymptotic or perturbative approximation. It also highlights the interplay between the effective parameter $E^2$ and the phase dynamics in determining whether a crossing can occur.
For example, for $\mathcal{E}>0$, the denominator in Eq.~\eqref{rho_c vs theta_c} must be positive, implying that the condition
\begin{equation}
2(\theta')^2>E^2
\end{equation}
must be satisfied at some stage of the evolution. While this condition is not sufficient to ensure the occurrence of a phantom crossing, it is necessary for its existence. Conversely, if
\begin{equation}
2(\theta')^2<E^2
\end{equation}
holds throughout the evolution, a phantom crossing is impossible. The possibility of a phantom crossing is therefore controlled by the interplay between the effective parameter $E^2$ and the phase dynamics. An analogous discussion applies for $\mathcal{E}<0$, with the corresponding inequalities reversed.

\begin{itemize}
\item {\bf $1-m^2/8<0$.}
For the oscillatory regime, Eq.~\eqref{analytic expansion} can be directly compared with the asymptotic perturbative expression derived in Eq.~\eqref{deltaw esplicita}. Indeed,

\begin{equation}
\delta w
=
\frac{1}{\rho^4}
\left[
\frac{2E^2}{y^2}
-
\frac{4(\theta')^2}{y^2}
+
\frac{6\mathcal E}{\rho^2y^2}
\right],
\end{equation}

while asymptotically

\begin{equation}
\delta w
=
\frac{1}{\rho^4}
\left[
\frac{2E^2}{\bar y^2}
+
h(\phi)
+
\frac{6C}{\rho^2}
\right].
\end{equation}

Although the latter has been obtained perturbatively, it reproduces the same qualitative structure as the exact expression, with the contribution from the phase dynamics encoded in the oscillatory function $h(\phi)$ and an additional subleading term proportional to the integration constant $C$. Neglecting this asymptotically vanishing contribution, the occurrence of a phantom crossing is determined, in both descriptions, by the competition between the term with the effective parameter $E^2$ and the phase dynamics, consistently with the discussion of the previous section.

\item {\bf $1-m^2/8>0$.}
For the non-oscillatory regime, Eq.~\eqref{zeta(rho)} gives
\begin{equation}
\theta'
=
-2\bar y
\left[
A(1+\mu)\rho^{-2\mu}
+
B(1-\mu)\rho^{2\mu}
\right],
\qquad
0<\mu<1.
\end{equation}
For a generic solution, the contribution proportional to $A$
decays asymptotically, whereas the one proportional to $B$ grows and therefore
provides the leading contribution at large $\rho$. Retaining only this
asymptotically dominant term, we obtain
\begin{equation}
\theta'
\simeq
-2\bar y(1-\mu)B\rho^{2\mu},
\end{equation}
and consequently
\begin{equation}
(\theta')^2
\sim
\rho^{4\mu}.
\end{equation}

Hence,

\begin{equation}
(\theta')^2
\gg
E^2,
\qquad
(\rho\rightarrow\infty),
\end{equation}

showing that the condition

\begin{equation}
2\theta'^2>E^2
\end{equation}

is always satisfied at sufficiently late times.

Consequently,

\begin{equation}
w+1
\simeq
-
\frac{4\theta'^2}
{\rho^4\bar y^2}
\sim
-\rho^{-4+4\mu},
\end{equation}

in agreement with the perturbative result

\begin{equation}
\delta w(\rho)
\sim
-\rho^{-4+4\mu}.
\end{equation}

Therefore, the equation of state parameter always approaches the de Sitter limit from the phantom side.\\

\item For completeness, the marginal case $1-m^2/8=0$ leads to
logarithmic corrections to the asymptotic behaviour,
$\delta w\sim-\log^2\rho/\rho^4$, and therefore still
approaches the de Sitter limit from the phantom side.
However, this case is not physically realized for the
allowed integer values of $m$, as already remarked.

\end{itemize}


\section{Dynamical dark energy with two-mode condensates and non-trivial phase dynamics}

So far, the analysis of the corrections to the asymptotic de Sitter regime
has been restricted to the single-mode sector, allowing us to isolate the
effects associated with the phase dynamics. We now return to the multimode
setting and retain, in addition to the asymptotically dominant mode, the
leading subdominant contribution, which we denote as the second condensate
mode. We restrict the analysis to the regime in which both modes have reached
the large amplitude asymptotic region, so that the asymptotic expansions
derived above can be consistently applied to each of them. 

For a two-mode condensate, the total volume is given by the sum of the
contributions of the two modes and Eq.~\eqref{V operator sigma} reduces to

\begin{equation}\label{two mode volume}
V(\chi_0)
=
V_1\rho_1^2(\chi_0)
+
V_2\rho_2^2(\chi_0).
\end{equation}

The corresponding effective equation-of-state parameter is
\begin{equation}
w_{\rm eff}
=
3-2\frac{VV''}{(V')^2}.
\end{equation}

To characterize the relative contribution of the two modes in the asymptotic
regime, we first consider the leading-order amplitudes $\rho_{j,0}$ obtained
in Eq.~\eqref{leading rho solution}. As discussed in Sec.~\ref{sec: Asymtotic dynamics}, we take
\begin{equation}
\chi_{0,1}^{(\infty)}
<
\chi_{0,2}^{(\infty)},
\end{equation}
so that the first mode reaches its asymptotic divergence before the second.
As $\chi_0\to\chi_{0,1}^{(\infty)}$, it then follows directly from
Eq.~\eqref{leading rho solution} that $\rho_{1,0}$ diverges, whereas
$\rho_{2,0}$ remains finite. Since we restrict the analysis to the regime in
which both modes have already reached the large-amplitude asymptotic region,
the latter remains large but finite.

We therefore introduce the relative contribution of the second mode to the
total volume,
\begin{equation}
r(\chi_0)
\equiv
\frac{V_2\rho_2^2}{V_1\rho_1^2},
\end{equation}
which satisfies $r\ll1$ sufficiently close to the asymptotic limit and
vanishes as $\chi_0\to\chi_{0,1}^{(\infty)}$. At leading order:
\begin{equation}
r_0(\chi_0)
=
\frac{V_2\rho_{2,0}^2}{V_1\rho_{1,0}^2}
=
\frac{V_2}{V_1}
\sqrt{\frac{|\lambda_1|}{|\lambda_2|}}
\frac{
\chi_{0,1}^{(\infty)}-\chi_0
}{
\chi_{0,2}^{(\infty)}-\chi_0
},
\end{equation}
which makes the limit $r_0\to0$ explicit.

The smallness of $r$ allows us to expand the effective equation of state
around the asymptotically dominant mode:
\begin{equation}\label{weff}
w_{\rm eff}
=
-1+\delta w_1
-r\left(4-\delta w_1\right)
+
2r^2
\frac{V_1y_2}{V_2y_1}
\left(4-\delta w_1\right)
+
\mathcal{O}(r^3),
\end{equation}
where $\delta w_1\equiv w_1+1$ denotes the single-mode correction associated
with the dominant mode, studied in the previous section.

The deviations from the leading two-mode behaviour depend on the phase
dynamics. As in the single-mode analysis, we therefore distinguish between
the oscillatory and non-oscillatory regimes.

\vspace{0.5cm}
{\bf Oscillatory regime}\label{subsec:oscillatory_regime}

In this interval admissible integer values of \(m\) exist, namely
\(m=-6,\pm4\).\\
In the oscillatory regime, the asymptotic result obtained in the single-mode
analysis can be applied independently to each condensate mode. We therefore
have
\begin{equation}\label{y_j asymptotic expansion}
y_j
=
\frac{\rho_j'}{\rho_j^3}
=
\bar y_j
\left[
1+
\frac{g_j(\phi_j)}{\rho_{j,0}^4}
+
\mathcal{O}(\rho_{j,0}^{-6})
\right],
\qquad
\phi_j=\beta\log\rho_{j,0},
\end{equation}

where \(g_j(\phi_j)\) is the periodic function given in
Eq.~\eqref{g oscillatory}.
In order to determine the corresponding correction to the contribution of
each mode to the total volume, we write the mode amplitudes as

\begin{equation}\label{rho_j asymptotic expansion}
\rho_j^2
=
\rho_{j,0}^2
\left[
1+
\frac{G_j(\phi_j)}{\rho_{j,0}^4}
+
\mathcal{O}(\rho_{j,0}^{-6})
\right].
\end{equation}
The function \(G_j(\phi_j)\) is fixed by requiring this expansion to reproduce
the asymptotic solution for \(y_j\). Indeed, differentiating
Eq.~\eqref{rho_j asymptotic expansion} with respect to \(\chi_0\) and using $(\rho_j^2)'
=
2y_j\rho_j^4$, together with the leading-order relation
\(\rho_{j,0}'=\bar y_j\rho_{j,0}^3\), gives

\begin{equation}\label{eq per g e G}
\frac{\beta}{2}
\frac{dG_j}{d\phi_j}
-
3G_j
=
g_j(\phi_j).
\end{equation}

Since \(g_j(\phi_j)\) contains a constant contribution together with the
harmonics \(\cos2\phi_j\) and \(\sin2\phi_j\), the solution has the same
harmonic structure,
\begin{equation}
G_j(\phi_j)
=
G_{0j}
+
G_{cj}\cos2\phi_j
+
G_{sj}\sin2\phi_j,
\end{equation}
with
\begin{equation}
G_{0j}
=
-\frac{E_j^2}{6\bar y_j^2}
+
\frac{k_{Aj}+k_{Bj}}{12},
\end{equation}
\begin{equation}
G_{cj}
=
\frac{
(\beta^2+3)(k_{Aj}-k_{Bj})
-
2\beta k_{ABj}
}{
4(1+\beta^2)(\beta^2+9)
},
\end{equation}
and
\begin{equation}
G_{sj}
=
\frac{
2\beta(k_{Aj}-k_{Bj})
+
(\beta^2+3)k_{ABj}
}{
4(1+\beta^2)(\beta^2+9)
}.
\end{equation}

The leading-order ratio between the two volume contributions is therefore
corrected according to

\begin{equation}
r
=
r_0
\left[
1+
\frac{G_2(\phi_2)}{\rho_{2,0}^4}
-
\frac{G_1(\phi_1)}{\rho_{1,0}^4}
\right].
\end{equation}

Substituting the asymptotic expansions of both modes into Eq.~\eqref{weff} and
ordering the resulting contributions according to the asymptotic limit

\[
\rho_{1,0}\rightarrow\infty,
\qquad
\rho_{2,0}\rightarrow\rho_{2,\infty}\gg1,
\]

yields

\begin{equation}\label{two mode oscillatory eos}
\begin{aligned}
w_{\rm eff}
=
-1
-
\underbrace{
4r_0
}_{\displaystyle
\mathcal{O}\!\left(
\frac{\rho_{2,0}^{2}}{\rho_{1,0}^{2}}
\right)}
-
\underbrace{
4r_0
\frac{G_2(\phi_2)}{\rho_{2,0}^{4}}
}_{\displaystyle
\mathcal{O}\!\left(
\frac{1}{\rho_{1,0}^{2}\rho_{2,0}^{2}}
\right)}
+
\underbrace{
8r_0^{2}
\frac{V_1}{V_2}
\sqrt{\frac{|\lambda_2|}{|\lambda_1|}}
}_{\displaystyle
\mathcal{O}\!\left(
\frac{\rho_{2,0}^{4}}{\rho_{1,0}^{4}}
\right)}
+
\underbrace{
\delta w_1
+
8r_0^{2}
\frac{V_1}{V_2}
\sqrt{\frac{|\lambda_2|}{|\lambda_1|}}
\frac{
g_2(\phi_2)+2G_2(\phi_2)
}{
\rho_{2,0}^{4}
}
}_{\displaystyle
\mathcal{O}\!\left(
\frac{1}{\rho_{1,0}^{4}}
\right)}
+
\cdots .
\end{aligned}
\end{equation}

Here

\begin{equation}
\delta w_1
=
\frac{1}{\rho_{1,0}^{4}}
\left[
\frac{2E_1^{2}}{\bar y_1^{2}}
+
h_1(\phi_1)
\right]
\end{equation}

is the leading correction generated by the contribution of the dominant mode to
the total volume.\\
Although the second mode remains subdominant in the total volume, its
phase-dependent contribution can exceed that of the dominant mode. This
behaviour is a direct consequence of the different asymptotic growth rates of
the two amplitudes. Since the dominant mode reaches the large-volume regime
first, its phase-dependent correction is suppressed as
\(\mathcal{O}(\rho_{1,0}^{-4})\). By contrast, the second mode does not diverge during
the physical evolution and its phase-dependent correction scales as
\(\mathcal{O}(\rho_{1,0}^{-2}\rho_{2,0}^{-2})\), remaining asymptotically larger as long
as \(\rho_{2,0}\) is finite.

Consequently, although the first mode dominates the total volume, the leading
deviations from the single-mode asymptotic equation of state are controlled by the
subdominant mode through both its direct contribution to the total volume and
its phase-dependent correction, which decays more slowly than the
phase-dependent correction associated with the dominant mode.


\vspace{0.5cm}

{\bf Non-oscillatory regime}

In the non-oscillatory regime, the leading perturbation of each mode behaves as

\begin{equation}
\xi_j(\rho_j)
\simeq
B_j\rho_j^{-2+2\mu},
\qquad
0<\mu<1.
\end{equation}

Using the leading-order solution introduced previously, the corresponding asymptotic expansion of \(y_j\) is

\begin{equation}\label{y_j nonosc asymptotic expansion}
y_j
=
\bar y_j
\left[
1+
m^2y^*B_j^2
\rho_j^{-4+4\mu}
+
\mathcal{O}(\rho_j^{-8+8\mu})
\right].
\end{equation}

The
condensate amplitudes admit the asymptotic expansion

\begin{equation}\label{rho_j nonosc asymptotic expansion}
\rho_j^2
=
\rho_{j,0}^2
\left[
1+
\frac{Q_j}{\rho_{j,0}^{4-4\mu}}
+
\mathcal{O}(\rho_{j,0}^{-8+8\mu})
\right],
\end{equation}

where

\begin{equation}
Q_j
=
\frac{4(1-\mu)}{3-2\mu}B_j^2.
\end{equation}

Accordingly,

\begin{equation}
r
=
r_0
\left[
1+
\frac{Q_2}{\rho_{2,0}^{4-4\mu}}
-
\frac{Q_1}{\rho_{1,0}^{4-4\mu}}
+
\mathcal{O}(\rho_{2,0}^{-8+8\mu})
\right].
\end{equation}

Here the leading phase-dependent correction associated with the dominant mode is given by

\begin{equation}
\delta w_1
=
-\frac{16(1-\mu)^2B_1^2}
{\rho_{1,0}^{4-4\mu}}.
\end{equation}

The resulting asymptotic hierarchy depends on the value of \(\mu\). 
However, the conditions imposed on the integer parameter $m$ restrict the analysis to $\frac12<\mu<1$, with $m=\pm2$ being the only admissible integer values.
In this interval the phase-dependent correction associated with the dominant mode becomes the leading asymptotic correction. The effective equation of state therefore reads

\begin{equation}
\begin{aligned}
w_{\rm eff}
=
-1
+
\underbrace{
\delta w_1
}_{\displaystyle
\mathcal{O}\!\left(
\frac{1}{\rho_{1,0}^{4-4\mu}}
\right)}
-
\underbrace{
4r_0
}_{\displaystyle
\mathcal{O}\!\left(
\frac{\rho_{2,0}^{2}}{\rho_{1,0}^{2}}
\right)}
-
\underbrace{
4r_0
\frac{Q_2}{\rho_{2,0}^{4-4\mu}}
}_{\displaystyle
\mathcal{O}\!\left(
\frac{\rho_{2,0}^{-2+4\mu}}
{\rho_{1,0}^{2}}
\right)}
+\cdots .
\end{aligned}
\end{equation}

The asymptotic dynamics of the
effective equation of state around \(w_{\rm eff}=-1\) is thus primarily
controlled by the phase-dependent correction of the dominant mode rather than
by the presence of the subdominant one.
\\

For completeness, in the marginal case $1-m^2/8=0$, the phase
perturbations acquire logarithmic corrections, with the leading correction
to the effective equation of state arising from the subdominant mode,
followed by its phase-dependent contribution and, at a further subleading
order, by the phase-dependent contribution of the dominant mode. This case,
however, is not realized for the admissible integer values of $m$, as already remarked.

\section{Effective equation of state in the phantom phase}
\label{sec:phantom_phase}

In the previous sections, we have characterized the approach to the
asymptotic de Sitter regime, first isolating the corrections induced by the
phase dynamics in the single-mode sector and then combining them with the
contribution of a second condensate mode. We now investigate more closely
the resulting evolution within the phantom phase.

In particular, we focus on the minimum reached by the effective equation of state before its asymptotic approach to $w_{\rm eff}=-1$. We characterize both the position and the depth of this minimum. Its position also provides a measure of the relational time separation between the strongest departure from the de Sitter value and the asymptotic regime.

The corresponding problem in the two-mode model with trivial phase
dynamics was studied in Ref.~\cite{Pang:2025jtk}. Here, we extend this
analysis by including the phase-dependent corrections derived in the
previous sections and determine how they modify the position and depth of
the phantom minimum.

As before, the form and asymptotic scaling of the phase corrections depend
on the interaction parameter $m$. We therefore analyze separately the
oscillatory and non-oscillatory regimes, which contain the physically
relevant cases considered in this work. The limiting case $\mu=0$, which is not realized by the allowed integer values of $m$, will not be considered in the following.

Starting from the two-mode volume in Eq. \eqref{two mode volume} the effective equation of state can be written as
\begin{equation}
\begin{aligned}
w_{\rm eff}
=
-1
-
\frac{
\left(V_1\rho_1^2+V_2\rho_2^2\right)
\left(V_1y_1'\rho_1^4+V_2y_2'\rho_2^4\right)
}{
\left(
V_1y_1\rho_1^4+V_2y_2\rho_2^4
\right)^2
}
-
4
\frac{
V_1V_2\rho_1^2\rho_2^2
\left(
y_1\rho_1^2-y_2\rho_2^2
\right)^2
}{
\left(
V_1y_1\rho_1^4+V_2y_2\rho_2^4
\right)^2
}.
\end{aligned}
\label{weff_minimum_analysis}
\end{equation}

This expression makes explicit the two effects discussed above. The first
correction contains the contribution associated with the phase dynamics,
through the evolution of the individual modes, whereas the second is the
genuine two-mode contribution already present in the absence of non-trivial phase dynamics. Since the phase-dependent corrections have different asymptotic behaviours in the oscillatory and non-oscillatory regimes, we analyze separately the two cases.


\vspace{0.5cm}

{\bf Oscillatory regime}

In the oscillatory regime, the asymptotic evolution of the two condensate
modes is described by the phase-dependent corrections derived in the
previous sections. In particular, the periodic function $g_j(\phi_j)$
entering the asymptotic evolution of $y_j$ is given in
Eq.~\eqref{g oscillatory}, while the corresponding correction
$G_j(\phi_j)$ to the condensate amplitude is defined through
Eq.~\eqref{rho_j asymptotic expansion}.

In the large-volume regime, these oscillatory contributions become
increasingly rapid as functions of relational time. Indeed, using the definition
$\phi_j=\beta\log\rho_{j,0}$, one finds
\begin{equation}
\phi_j'
\simeq
\beta\bar y_j\rho_{j,0}^2,
\end{equation}
so that the oscillation frequency increases with the condensate amplitude,
while their amplitude remains perturbatively suppressed.

Keeping the full oscillatory dependence would make the analytical
determination of the minimum of $w_{\rm eff}$ from
Eq.~\eqref{weff_minimum_analysis} considerably more involved. Moreover,
these increasingly rapid oscillations represent a small modulation around a
more slowly varying profile of the effective equation of state. Since our
aim is to characterize the overall location and depth of the phantom
minimum, rather than the sequence of local extrema generated by this rapid
modulation, we coarse grain over the oscillatory contributions and focus on
the smooth profile around which the full effective equation of state
oscillates. In this way, the rapid and perturbatively small modulation is
averaged out, while the non-oscillatory contribution of the phase dynamics
is retained.

Accordingly,
\begin{equation}
\left\langle\cos 2\phi_j\right\rangle
=
\left\langle\sin 2\phi_j\right\rangle
=
0,
\end{equation}
so that
\begin{equation}
G_j(\phi_j)\longrightarrow G_{0j},
\qquad
g_j(\phi_j)\longrightarrow g_{0j}.
\end{equation}

Taking the non-oscillatory part of Eq.~\eqref{eq per g e G} then gives
\begin{equation}
g_{0j}
=
-3G_{0j}.
\label{g0G0relation}
\end{equation}

Applying the coarse-graining prescription to
Eqs.~\eqref{rho_j asymptotic expansion} and
\eqref{y_j asymptotic expansion}, and using the leading-order solution
in Eq.~\eqref{leading rho solution}, we obtain
\begin{equation}
\rho_j^2(\chi_0)
=
\frac{1}{
2\bar y_j
\left(
\chi_{0,j}^{(\infty)}-\chi_0
\right)}
+
2\bar y_jG_{0j}
\left(
\chi_{0,j}^{(\infty)}-\chi_0
\right),
\label{rho_coarse_grained}
\end{equation}
and
\begin{equation}
y_j(\chi_0)
=
\bar y_j
\left[
1+
4\bar y_j^2g_{0j}
\left(
\chi_{0,j}^{(\infty)}-\chi_0
\right)^2
\right].
\label{y_coarse_grained}
\end{equation}


\begin{itemize}

\item {\bf Location of the minimum}

Following the same large-volume strategy used in
Ref.~\cite{Pang:2025jtk}, we assume that the minimum of the effective
equation of state is reached close to
$\chi_{0,1}^{(\infty)}$. We therefore write
\begin{equation}
\chi_0
=
\chi_{0,1}^{(\infty)}-\delta,
\qquad
\delta
\ll
\chi_{0,2}^{(\infty)}
-
\chi_{0,1}^{(\infty)},
\label{delta_definition}
\end{equation}
where $\delta$ measures the relational time separation between the minimum
and the asymptotic divergence of the dominant mode.

We then substitute the coarse-grained solutions
\eqref{rho_coarse_grained} and \eqref{y_coarse_grained} into
Eq.~\eqref{weff_minimum_analysis} and impose the minimum condition
$w_{\rm eff}'=0$. Expanding around $\delta=0$ and consistently retaining
the leading contribution in $\delta$ and terms up to first order in
$G_{01}$ and $G_{02}$, we obtain
\begin{equation}
\begin{aligned}
\delta
\simeq
\frac{
\chi_{0,2}^{(\infty)}
-
\chi_{0,1}^{(\infty)}
}{6}
\Bigg[
1
&+
4
\left(
\chi_{0,2}^{(\infty)}
-
\chi_{0,1}^{(\infty)}
\right)^2
\frac{V_1}{V_2}
\bar y_1\bar y_2G_{01}
+
8
\left(
\chi_{0,2}^{(\infty)}
-
\chi_{0,1}^{(\infty)}
\right)^2
\bar y_2^2G_{02}
\Bigg].
\end{aligned}
\label{delta_min_phase}
\end{equation}

This result separates naturally into the phase-free two-mode contribution
and the correction induced by the phase dynamics. Writing
\begin{equation}
\delta
=
\delta_0+\Delta\delta^{\rm phase},
\end{equation}
the phase-free contribution is
\begin{equation}
\delta_0
=
\frac{
\chi_{0,2}^{(\infty)}
-
\chi_{0,1}^{(\infty)}
}{6},
\label{delta_no_phase}
\end{equation}
while the phase-dependent correction is
\begin{equation}
\begin{aligned}
\Delta\delta^{\rm phase}
={}&
\frac{2}{3}
\left(
\chi_{0,2}^{(\infty)}
-
\chi_{0,1}^{(\infty)}
\right)^3
\frac{V_1}{V_2}
\bar y_1\bar y_2G_{01}
+
\frac{4}{3}
\left(
\chi_{0,2}^{(\infty)}
-
\chi_{0,1}^{(\infty)}
\right)^3
\bar y_2^2G_{02}.
\end{aligned}
\label{phase_correction_delta}
\end{equation}

Equation~\eqref{delta_no_phase} reproduces the large-volume result obtained
in Ref.~\cite{Pang:2025jtk} for the two-mode model without phase dynamics.
Indeed, switching off the phase corrections,
$G_{01}=G_{02}=0$, directly gives
$\Delta\delta^{\rm phase}=0$.

In general, the sign of $g_{0j}$, and therefore that of $G_{0j}$, is not
fixed a priori. As discussed in Sec.~\ref{oscillatory regime in Sec 4}, for a
single condensate mode the requirement of a phantom-like phase correction
constrains the contribution of the second scalar field, suggesting an upper
bound on its conserved momentum $\bar\pi_\eta$.

If we assume that this condition is satisfied not only for the dominant
mode, but also for the second condensate mode, the coarse-grained phase
correction is phantom-like for both modes. We then have
$g_{01}<0$ and $g_{02}<0$, and, using Eq.~\eqref{g0G0relation},
$G_{01}>0$ and $G_{02}>0$. Under this assumption,
Eq.~\eqref{phase_correction_delta} implies
\begin{equation}
\Delta\delta^{\rm phase}>0.
\end{equation}

Since $\chi_{0,\min}=\chi_{0,1}^{(\infty)}-\delta$, the phase dynamics therefore shifts the phantom minimum toward earlier relational times with respect to the phase-free two-mode evolution. Equivalently, it increases the relational time separation between the phantom minimum and the asymptotic regime in which $w_{\rm eff}$ approaches $-1$.


\item {\bf Minimum value of the effective equation of state}

Having determined the position of the minimum, we now evaluate the
coarse-grained effective equation of state at
$\chi_{0,\min}=\chi_{0,1}^{(\infty)}-\delta$, with $\delta$ given by
Eq.~\eqref{delta_min_phase}. Substituting the coarse-grained solutions into
Eq.~\eqref{weff_minimum_analysis} and consistently retaining terms up to
first order in $G_{01}$ and $G_{02}$, the result can be conveniently written as
\begin{equation}
w_{\min}
=
w_{\min}^{\rm no\,phase}
+
\Delta w_{\min}^{\rm phase}.
\label{wmin_decomposition}
\end{equation}

The first term is the contribution already present in the phase-free
two-mode model,
\begin{equation}
w_{\min}^{\rm no\,phase}
=
-1
-
\frac{
1008V_1V_2\bar y_1\bar y_2
}{
\left(
49V_1\bar y_2+V_2\bar y_1
\right)^2
},
\label{wmin_no_phase}
\end{equation}
which reproduces the large volume result of
Ref.~\cite{Pang:2025jtk}. Thus, switching off the phase dynamics recovers
the previous two-mode result, as expected.

The remaining contribution contains the effect of the coarse-grained phase
dynamics:
\begin{equation}
\begin{aligned}
\Delta w_{\min}^{\rm phase}
={}&
-
\frac{4\bar y_1\bar y_2}{
3\left(49V_1\bar y_2+V_2\bar y_1\right)^3
}
\left(
\chi_{0,2}^{(\infty)}-\chi_{0,1}^{(\infty)}
\right)^2
\\
&\times
\Bigg\{
G_{01}
\Big[
202321V_1^3\bar y_1\bar y_2^2
+
3708V_1^2V_2\bar y_1^2\bar y_2
+
35V_1V_2^2\bar y_1^3
\Big]
\\
&\qquad
+
G_{02}
\Big[
522291V_1^2V_2\bar y_2^3
+
556V_1V_2^2\bar y_1\bar y_2^2
+
49V_2^3\bar y_1^2\bar y_2
\Big]
\Bigg\}.
\end{aligned}
\label{phase_correction_wmin}
\end{equation}

As discussed above, the signs of $G_{01}$ and $G_{02}$ are not fixed a
priori. However, if we restrict to the regime in which the condition for a
phantom-like coarse-grained phase correction is satisfied by both
condensate modes, then $G_{01}>0$ and $G_{02}>0$. Under this assumption,
all the terms in parentheses in Eq.~\eqref{phase_correction_wmin} are
positive, while the overall prefactor is negative. Therefore,
\begin{equation}
\Delta w_{\min}^{\rm phase}<0,
\qquad
w_{\min}<w_{\min}^{\rm no\,phase}.
\label{phase_deeper_minimum}
\end{equation}

The coarse-grained phase dynamics therefore makes the phantom minimum deeper
than in the corresponding phase-free two-mode evolution, provided that the
phantom-like phase condition is satisfied for both modes.

Together with the result for the location of the minimum, this shows a
twofold effect of the phase dynamics in this regime. The positive correction
$\Delta\delta^{\rm phase}$ increases the relational-time separation between
the minimum and $\chi_{0,1}^{(\infty)}$, shifting the minimum toward earlier
relational times, while the negative correction
$\Delta w_{\min}^{\rm phase}$ makes its value more phantom-like. Hence,
under the assumptions specified above, the phase dynamics both anticipates
and deepens the phantom minimum with respect to the phase-free two-mode
model of Ref.~\cite{Pang:2025jtk}.

\end{itemize}


\vspace{0.5cm}

{\bf Non-oscillatory regime: $0<\mu<1$}
\label{subsec:minimum_nonoscillatory}

We now extend the analysis of the phantom minimum to the non-oscillatory
regime. In contrast with the oscillatory case, no coarse-graining procedure
is required, since the phase-dependent corrections do not oscillate.

The large volume solutions for the two condensate modes have already been
derived in Eqs.~\eqref{y_j nonosc asymptotic expansion} and
\eqref{rho_j nonosc asymptotic expansion}. Using the leading-order relation
\eqref{leading rho solution} and the relation $m^2=8(1-\mu^2)$, these solutions can be expressed directly in
terms of the relational time as

\begin{equation}
y_j(\chi_0)
=
\bar y_j
\Biggl[
1
-
4(1-\mu)B_j^2
\left[
2\bar y_j
\left(
\chi_{0,j}^{(\infty)}-\chi_0
\right)
\right]^{2(1-\mu)}
\Biggr]
\label{yj_nonosc_minimum}
\end{equation}

and

\begin{equation}
\rho_j^2(\chi_0)
=
\frac{1}{
2\bar y_j
\left(
\chi_{0,j}^{(\infty)}-\chi_0
\right)}
\Biggl[
1
+
\frac{4(1-\mu)}{3-2\mu}B_j^2
\left[
2\bar y_j
\left(
\chi_{0,j}^{(\infty)}-\chi_0
\right)
\right]^{2(1-\mu)}
\Biggr].
\label{rhoj_nonosc_minimum}
\end{equation}

These expressions include the leading phase-dependent corrections to the
asymptotic evolution of each mode. Since $0<\mu<1$, these corrections vanish
as $\chi_0\to\chi_{0,j}^{(\infty)}$, consistently with the recovery of the
leading asymptotic solution.

We can therefore use Eqs.~\eqref{yj_nonosc_minimum} and \eqref{rhoj_nonosc_minimum} in
Eq.~\eqref{weff_minimum_analysis} to determine how the phase
dynamics modifies the position and the value of the phantom minimum.

\begin{itemize}

\item {\bf Location of the minimum}

As in the oscillatory regime, we consider the minimum of the effective
equation of state located close to $\chi_{0,1}^{(\infty)}$ and parametrize
its position as
\begin{equation}
\chi_0
=
\chi_{0,1}^{(\infty)}-\delta,
\qquad
\delta
\ll
\chi_{0,2}^{(\infty)}
-
\chi_{0,1}^{(\infty)}.
\label{delta_definition_nonosc}
\end{equation}

For generic $\mu$, substituting the non-oscillatory solutions
for $y_j$ and $\rho_j$ into the effective equation of state produces
non-integer powers, making a direct analytic solution of the stationarity
condition $w_{\rm eff}'=0$ cumbersome. We therefore exploit the perturbative
character of the phase corrections and determine the position of the minimum
as a perturbation of the phase-free result.

In the absence of phase dynamics, the position of the minimum is given by
Eq.~\eqref{delta_no_phase}, namely
\begin{equation}
\delta_0
=
\frac{
\chi_{0,2}^{(\infty)}
-
\chi_{0,1}^{(\infty)}
}{6},
\end{equation}
where $\delta_0$ denotes the phase-free distance of the minimum from
$\chi_{0,1}^{(\infty)}$.

To determine the correction induced by the phase dynamics, we write the
effective equation of state as
$w_{\rm eff}=w_0+w_{\rm phase}+\mathcal{O}(B_j^4)$, where $w_0$ denotes
the phase-free contribution and $w_{\rm phase}$ the leading
phase-dependent correction, of order $\mathcal{O}(B_j^2)$. Correspondingly,
the position of the minimum is shifted from $\delta_0$ by a quantity
$\Delta\delta^{\rm phase}=\mathcal{O}(B_j^2)$.

Expanding the stationarity condition to first order around the phase-free
minimum gives
\begin{equation}
\Delta\delta^{\rm phase}
=
-
\frac{
w_{\rm phase}'(\delta_0)
}{
\left.
\dfrac{\partial w_0'}{\partial\delta}
\right|_{\delta_0}
}
=
B_1^2\mathcal{D}_1(\mu)
+
B_2^2\mathcal{D}_2(\mu).
\label{delta_phase_nonosc}
\end{equation}
Here, $\mathcal{D}_1(\mu)$ and $\mathcal{D}_2(\mu)$ are the coefficients
encoding the displacement of the minimum induced by the phase dynamics of
the first and second condensate modes, respectively.

The position of the minimum, including the leading phase-dependent
correction, is therefore
\begin{equation}
\delta(\mu)
=
\frac{
\chi_{0,2}^{(\infty)}
-
\chi_{0,1}^{(\infty)}
}{6}
+
B_1^2\mathcal{D}_1(\mu)
+
B_2^2\mathcal{D}_2(\mu).
\label{delta_nonosc_final}
\end{equation}

The coefficient $\mathcal{D}_1(\mu)$, describing the displacement generated
by the phase dynamics of the first condensate mode, is
\begin{equation}
\begin{aligned}
\mathcal{D}_1(\mu)
={}&
-
\frac{
8\,3^{2\mu}
\left(\chi_{0,2}^{(\infty)}-\chi_{0,1}^{(\infty)}\right)^{3-2\mu}
V_1^2
\bar y_1^{\,1-2\mu}
\bar y_2^2
(\mu-1)
}{
9V_2(2\mu-3)
\left(
49V_1\bar y_2+V_2\bar y_1
\right)^4
}
\,
\mathcal{P}_1(\mu),
\end{aligned}
\label{D1_nonosc}
\end{equation}
where $\mathcal{P}_1(\mu)$ is the polynomial function
\begin{equation}
\begin{aligned}
\mathcal{P}_1(\mu)
={}&
V_1^3\bar y_2^3
\left(
11529602\mu^3
-40353607\mu^2
+46118408\mu
-17294403
\right)
\\
&+
V_1^2V_2\bar y_1\bar y_2^2
\left(
2117682\mu^3
-7311045\mu^2
+7498323\mu
-2132088
\right)
\\
&+
V_1V_2^2\bar y_1^2\bar y_2
\left(
72030\mu^3
-149205\mu^2
+8526\mu
+26313
\right)
\\
&+
V_2^3\bar y_1^3
\left(
686\mu^3
-343\mu^2
+35\mu
-18
\right).
\end{aligned}
\label{P1_nonosc}
\end{equation}

Similarly, the coefficient $\mathcal{D}_2(\mu)$, describing the displacement
generated by the phase dynamics of the second condensate mode, is
\begin{equation}
\begin{aligned}
\mathcal{D}_2(\mu)
={}&
-
\frac{
56\,3^{2\mu}
\left(\chi_{0,2}^{(\infty)}-\chi_{0,1}^{(\infty)}\right)^{3-2\mu}
V_1
\bar y_2^{\,3-2\mu}
(\mu-1)
}{
9\,49^\mu
(2\mu-3)
\left(
49V_1\bar y_2+V_2\bar y_1
\right)^4
}
\,
\mathcal{P}_2(\mu),
\end{aligned}
\label{D2_nonosc}
\end{equation}
where $\mathcal{P}_2(\mu)$ is the polynomial function
\begin{equation}
\begin{aligned}
\mathcal{P}_2(\mu)
={}&
V_1^3\bar y_2^3
\left(
33614\mu^3
-823543\mu^2
+4521083\mu
-4941258
\right)
\\
&+
V_1^2V_2\bar y_1\bar y_2^2
\left(
6174\mu^3
-149205\mu^2
+697662\mu
-258279
\right)
\\
&+
V_1V_2^2\bar y_1^2\bar y_2
\left(
210\mu^3
-3045\mu^2
-5397\mu
+5712
\right)
\\
&+
V_2^3\bar y_1^3
\left(
2\mu^3
-7\mu^2
+8\mu
-3
\right).
\end{aligned}
\label{P2_nonosc}
\end{equation}

Unlike in the oscillatory regime, the phase-induced displacement does not
have a definite sign. Indeed, the signs of $\mathcal{D}_1(\mu)$ and
$\mathcal{D}_2(\mu)$ depend on $\mu$ and on the remaining model parameters
through the functions $\mathcal{P}_1(\mu)$ and $\mathcal{P}_2(\mu)$.
Consequently, the total correction $\Delta\delta^{\rm phase}$ can be either positive or negative. The phase dynamics can thus shift the phantom minimum toward earlier or later relational times, depending on the specific parameters of the model.


\item {\bf Minimum value of the effective equation of state}
\label{subsubsec:wmin_nonosc}

We finally determine the value of the effective equation of state at the
minimum in the non-oscillatory regime. As in the analysis of its location,
we separate the effective equation of state into the phase-free contribution
$w_0$ and the leading phase-dependent correction $w_{\rm phase}$, of order
$\mathcal{O}(B_j^2)$.

The minimum is located at
$\delta=\delta_0+\Delta\delta^{\rm phase}$, with
$\Delta\delta^{\rm phase}=\mathcal{O}(B_j^2)$, as derived in
Eq.~\eqref{delta_phase_nonosc}. Expanding the effective equation of state
around the phase-free minimum $\delta_0$, the term proportional to
$\Delta\delta^{\rm phase}$ vanishes at first order because
$\delta_0$ is a stationary point of $w_0$. Moreover, the variation of
$w_{\rm phase}$ induced by the displacement of the minimum contributes only at order $\mathcal{O}(B_j^4)$. Therefore, although the phase dynamics shifts the position of the minimum, this displacement does not affect its depth at the perturbative order considered here.

The minimum value is consequently obtained by evaluating the leading
phase-dependent correction at the phase-free minimum. Using the phase-free
result in Eq.~\eqref{wmin_no_phase}, we obtain
\begin{equation}
\begin{aligned}
w_{\min}
\simeq
-1
&-
\frac{
1008V_1V_2\bar y_1\bar y_2
}{
\left(
49V_1\bar y_2
+
V_2\bar y_1
\right)^2
}
+
B_1^2\mathcal{W}_1(\mu)
+
B_2^2\mathcal{W}_2(\mu).
\end{aligned}
\label{wmin_nonosc}
\end{equation}
Here, $\mathcal{W}_1(\mu)$ and $\mathcal{W}_2(\mu)$ are the coefficients
encoding the leading corrections to the value of the minimum generated by
the phase dynamics of the first and second condensate modes.

The coefficient associated with the phase dynamics of the first mode is
\begin{equation}
\begin{aligned}
\mathcal{W}_1(\mu)
={}&
-
\frac{
112\,3^{2\mu}(1-\mu)
V_1\bar y_2
\left[
\left(
\chi_{0,2}^{(\infty)}
-
\chi_{0,1}^{(\infty)}
\right)
\bar y_1
\right]^{2-2\mu}
}{
9(3-2\mu)
\left(
49V_1\bar y_2
+
V_2\bar y_1
\right)^3
}
\\
&\times
\Bigg[
V_1^2\bar y_2^2
\left(
33614\mu^2
-
84035\mu
+
50421
\right)
\\
&\qquad
+
V_1V_2\bar y_1\bar y_2
\left(
5488\mu^2
-
12544\mu
+
5292
\right)
\\
&\qquad
+
V_2^2\bar y_1^2
\left(
98\mu^2
-
77\mu
+
15
\right)
\Bigg].
\end{aligned}
\label{W1_nonosc}
\end{equation}

The corresponding coefficient associated with the phase dynamics of the
second mode is
\begin{equation}
\begin{aligned}
\mathcal{W}_2(\mu)
={}&
-
\frac{
784\,3^{2\mu}(1-\mu)
V_2\bar y_1
\left[
\left(
\chi_{0,2}^{(\infty)}
-
\chi_{0,1}^{(\infty)}
\right)
\bar y_2
\right]^{2-2\mu}
}{
9\,7^{2\mu}(3-2\mu)
\left(
49V_1\bar y_2
+
V_2\bar y_1
\right)^3
}
\\
&\times
\Bigg[
V_1^2\bar y_2^2
\left(
686\mu^2
-
9947\mu
+
21609
\right)
\\
&\qquad
+
V_1V_2\bar y_1\bar y_2
\left(
112\mu^2
-
1456\mu
+
1092
\right)
\\
&\qquad
+
V_2^2\bar y_1^2
\left(
2\mu^2
-
5\mu
+
3
\right)
\Bigg].
\end{aligned}
\label{W2_nonosc}
\end{equation}

The signs of these two phase corrections can be determined by analysing the
quadratic polynomials in $\mu$ appearing in
Eqs.~\eqref{W1_nonosc} and \eqref{W2_nonosc}. For the first mode,
$\mathcal{W}_1(\mu)$ is negative independently of the remaining model
parameters for
$0<\mu\leq5/14$ and $3/7\leq\mu<\mu_{c,1}$, where
$\mu_{c,1}\simeq0.6412717227$. Outside these intervals, its sign depends on
the model parameters. For the second mode,
$\mathcal{W}_2(\mu)$ is negative for
$0<\mu<\mu_{c,2}$, with $\mu_{c,2}\simeq0.8723218897$, while above this
critical value its sign becomes parameter dependent.

For the physically relevant non-oscillatory case $m=\pm2$, i.e. $\mu=1/\sqrt{2}\simeq0.7071$, the second-mode coefficient is always negative, $\mathcal{W}_2(1/\sqrt{2})<0$, whereas the sign of $\mathcal{W}_1(1/\sqrt{2})$ depends on the model parameters. Hence the phase dynamics of the second mode always deepens the phantom minimum, while that of the first mode can either deepen or raise it. The sign of the total phase-dependent correction $B_1^2\mathcal{W}_1+B_2^2\mathcal{W}_2$ is therefore not fixed a priori.

\end{itemize}

Overall, the phase dynamics modifies both the location and the depth of the
phantom minimum with respect to the evolution generated by the two condensate
amplitudes alone. The nature of this modification depends on the asymptotic
behaviour of the phase perturbations. In the oscillatory regime, after
coarse-graining over the rapid oscillations, the phase contribution shifts
the minimum toward earlier relational times and makes it deeper. In the
non-oscillatory regime, instead, neither effect is universal: both the
displacement and the correction to the depth generally depend on the model
parameters. In particular, for the physically relevant case $m=\pm2$, the
phase dynamics of the second mode always deepens the minimum, whereas the
contribution of the first mode is parameter dependent, so that the net
phase-induced correction to its depth is not fixed a priori.


\section{Conclusions}
We have analysed the late-time cosmological dynamics emerging from an interacting tensorial group field theory model, generalizing and combining recent analyses in this quantum gravity formalism, specifically those in \cite{Marchetti:2025jze} and in \cite{Pang:2025jtk, Oriti:2021rvm}, in a mean field approximation and under a relational definition of the cosmological evolution. We considered a larger parameter set, removing some simplifying assumptions adopted in previous work, and we included an additional scalar matter field, which plays the role of a non-trivial matter content of the universe, besides the scalar matter used as a clock; more importantly, we considered both the impact from two condensate modes and from the condensate phase, thus combining the two mechanisms for producing a dynamical dark energy identified in earlier work, and then analyzing their respective contribution.

We have confirmed that the fundamental quantum gravity interactions (of order six) produce, at an effective hydrodynamic approximation, a cosmological acceleration that could be then re-interpreted as a dynamical dark energy term, but without invoking new exotic matter fields, and only relying on the fundamental quantum gravity dynamics, depending on a very limited set of parameters characterizing it and the choice of condensate states. In a large range of parameter space, this emergent dark energy component has a phantom-like behaviour.

The results of our analysis, therefore, increase the robustness and generality of the suggested TGFT explanation for (dynamical) dark energy. The resulting cosmological dynamics from earlier work has been already compared with the most recent cosmological observations in \cite{LUCA-DESI}, confirming its promise, so our analysis can make the connection with cosmological observations, and thus the proposed explanatory mechanism for dark energy, even more robust and compelling. 

A first step is in fact to repeat the kind of analysis done in \cite{LUCA-DESI} for the more general model we have studied in this contribution.

There are also many directions for improving the class of models we have analyzed. 

First of all, we have studied the fundamental quantum gravity dynamics in the simplest mean field approximation, and then further restricted it to isotropic degrees of freedom only. While the latter restriction is customary in theoretical cosmology (for large-scale dynamics), it is certainly one that can and should be lifted \cite{Aniso, TGFTcosmologyANISOTROPIES}. The former approximation should also be lifted, to check how quantum corrections modify the emergent cosmological dynamics, either by studying one-loop corrections to the TGFT mean field action using TGFT renormalization techniques \cite{TGFTrenorm} or developing further the recent work on collective quantum excitations over TGFT condensates via Bogolyubov theory \cite{GFTbogolons}. 

Even remaining at the mean field level, we could improve our reference models by considering the coupled dynamics of cosmological perturbations \cite{TGFTpertu} and their backreaction on the cosmological evolution, and by making them more realistic in their matter content, going beyond the simple (relativistic, free) scalar fields considered so far, while at the same time strengthening their connection with the most promising TGFT models for 4d quantum gravity \cite{TGFTs, SF, GFT-LQG, TGFT-SF}.

Still, the very fact that such quantum gravity models provide a novel and rather robust mechanism for explaining a dynamical dark energy in terms of their fundamental dynamics, with few parameters to adjust and allowing a direct comparison with cosmological observations, is exciting and promising.

\section*{Acknowledgements}
We thank Andrea Calcinari, Luca Marchetti and Mariaveronica De Angelis for discussions and useful comments on our work. DO acknowledges support from Grant PR28/23 ATR2023-145735 funded by the Agencia Estatal de Investigación of the Spanish Government through \newline 
MCIN/AEI/10.13039/501100011033.
The authors further acknowledge support from the WOST (WithOut SpaceTime) project, supported by Grant ID 63683 from the John Templeton Foundation.


\newpage

\appendix

\section{Asymptotic corrections to the single-mode dynamics}
\label{app:single-mode-corrections}

In this appendix, we provide the technical details of the asymptotic
expansions used in Sec.~\ref{sec:dynamical-dark-energy} to determine the
leading corrections to the effective equation of state in the single-mode
sector. We first consider the oscillatory regime and then the
non-oscillatory one.

\subsection{Oscillatory regime}
\label{app:oscillatory-regime}

We consider the oscillatory regime $1-\frac{m^2}{8}<0$
and specialize the dynamical system in Eq.~\eqref{x', y', z'} to $l=5$.
The equation governing $y$ is then
\begin{equation}
    y'
    =
    -3\rho^2y^2
    +\frac{z^2+E^2}{\rho^2}
    +\frac{12\lambda}{6-m}\rho^2
    \cos(\varphi+mx).
\end{equation}

We expand the dynamical variables around the asymptotic fixed point as
\begin{equation}
    x=\bar{x}+\xi,
    \qquad
    y=\bar{y}+\iota,
    \qquad
    z=\zeta,
\end{equation}
where $\xi$, $\iota$ and $\zeta$ denote small perturbations. Expanding the
evolution equation for $y$ up to second order in the perturbations and using
Eq.~\eqref{y^bar^2}, we obtain
\begin{equation}
    \iota'
    =
    -6\rho^2\bar{y}\iota
    +\frac{\zeta^2+E^2}{\rho^2}
    -\frac{3}{2}m^2\bar{y}^2\rho^2\xi^2.
    \label{app:iota-osc}
\end{equation}
The quadratic contribution proportional to $\iota^2$ is subleading in the
asymptotic regime and can therefore be consistently neglected.

Using the asymptotic solutions for $\xi$ and $\zeta$ given in
Eqs.~\eqref{xi(rho)} and \eqref{zeta(rho)}, Eq.~\eqref{app:iota-osc} becomes
\begin{equation}
    \iota'
    =
    -6\rho^2\bar{y}\iota
    +\frac{\bar{y}^2}{\rho^2}F_E(\phi),
    \label{app:iota-FE}
\end{equation}
where
\begin{align}
    F_E(\phi)
    :=\;&
    \frac{E^2}{\bar{y}^2}
    +\left(
        4c_A^2-\frac{3}{2}m^2A^2
    \right)\cos^2\phi
    +
    \left(
        4c_B^2-\frac{3}{2}m^2B^2
    \right)\sin^2\phi
    +
    \left(
        8c_Ac_B-3m^2AB
    \right)\sin\phi\cos\phi .
    \label{app:FE}
\end{align}

To express the evolution equation in terms of $\rho$, we use $\iota'=\rho'\dfrac{d\iota}{d\rho}$,
together with the leading asymptotic relation $\rho'\simeq\bar{y}\rho^3$. Equation~\eqref{app:iota-FE} therefore reduces to
\begin{equation}
    \frac{d\iota}{d\rho}
    +\frac{6}{\rho}\iota
    =
    \frac{\bar{y}}{\rho^5}F_E(\phi).
    \label{app:iota-rho}
\end{equation}

The homogeneous solution is
\begin{equation}
    \iota_{\mathrm{hom}}
    =
    \frac{C}{\rho^6},
\end{equation}
in agreement with the linear result in
Eq.~\eqref{iota linear solution}. For the particular solution, the
asymptotic structure of the source term motivates the ansatz
\begin{equation}
    \iota_{\mathrm{part}}
    =
    \bar{y}\frac{g(\phi)}{\rho^4}.
    \label{app:iota-part}
\end{equation}
Since
\begin{equation}
    \phi=\beta\log\rho,
    \qquad
    \beta=2\sqrt{-1+\frac{m^2}{8}},
\end{equation}
substitution of Eq.~\eqref{app:iota-part} into
Eq.~\eqref{app:iota-rho} gives
\begin{equation}
    \beta\frac{dg(\phi)}{d\phi}
    +2g(\phi)
    =
    F_E(\phi).
    \label{app:g-equation}
\end{equation}

The solution is
\begin{align}
    g(\phi)
    =\;&
    \frac{E^2}{2\bar{y}^2}
    -\frac{k_A+k_B}{4} +
    \frac{-k_A+k_B+\beta k_{AB}}
    {4(1+\beta^2)}
    \cos 2\phi+
    \frac{-\beta k_A+\beta k_B-k_{AB}}
    {4(1+\beta^2)}
    \sin 2\phi ,
    \label{app:g-solution}
\end{align}
where
\begin{equation}
    k_A
    \equiv
    \frac{3}{2}m^2A^2-4c_A^2,
    \qquad
    k_B
    \equiv
    \frac{3}{2}m^2B^2-4c_B^2,
    \qquad
    k_{AB}
    \equiv
    3m^2AB-8c_Ac_B.
\end{equation}

Combining the homogeneous and particular solutions, and rescaling the
integration constant according to $C\rightarrow\bar{y}C$, the asymptotic
solution for $y$ can be written as
\begin{equation}
    y
    =
    \bar{y}
    \left(
        1+\frac{g(\phi)}{\rho^4}
        +\frac{C}{\rho^6}
    \right).
    \label{app:y-osc}
\end{equation}

Its derivative is, at the relevant asymptotic order,
\begin{equation}
    y'
    =
    \frac{\bar{y}^2}{\rho^2}
    \left(
        \beta\frac{dg(\phi)}{d\phi}
        -4g(\phi)
        -\frac{6C}{\rho^2}
    \right).
    \label{app:yprime-osc}
\end{equation}
Using Eq.~\eqref{app:g-solution}, this gives
\begin{equation}
    \frac{y'}{\rho^2\bar{y}^2}
    =
    \frac{1}{\rho^4}
    \left(
        -\frac{2E^2}{\bar{y}^2}
        +k_A+k_B
        +\tilde{k}_1\cos2\phi
        +\tilde{k}_2\sin2\phi
        -\frac{6C}{\rho^2}
    \right),
    \label{app:yprime-expanded}
\end{equation}
where
\begin{equation}
    \tilde{k}_1
    \equiv
    \frac{
        (2-\beta^2)(k_A-k_B)-3\beta k_{AB}
    }{
        2(1+\beta^2)
    },
\end{equation}
and
\begin{equation}
    \tilde{k}_2
    \equiv
    \frac{
        3\beta(k_A-k_B)+(2-\beta^2)k_{AB}
    }{
        2(1+\beta^2)
    }.
\end{equation}

Finally, using the single-mode equation of state in
Eq.~\eqref{deltaw(rho)}, we obtain

\begin{equation}
    \delta w(\rho)
    =
    \frac{1}{\rho^4}
    \left[
        \frac{2E^2}{\bar{y}^2}
        +\delta w_0
        +\delta w_1\cos2\phi
        +\delta w_2\sin2\phi
        +\frac{6C}{\rho^2}
    \right],
    \label{app:deltaw-osc}
\end{equation}

where

\begin{equation}
    \delta w_0\equiv-(k_A+k_B),
    \qquad
    \delta w_1\equiv-\tilde{k}_1,
    \qquad
    \delta w_2\equiv-\tilde{k}_2.
\end{equation}

Defining the oscillatory contribution
\begin{equation}
    h(\phi)
    =
    \delta w_0
    +\delta w_1\cos2\phi
    +\delta w_2\sin2\phi,
    \label{app:h-phi}
\end{equation}
one finds the equation

\begin{equation}
    \delta w(\rho)
    =
    \frac{1}{\rho^4}
    \left[
        \frac{2E^2}{\bar{y}^2}
        +h(\phi)
        +\frac{6C}{\rho^2}
    \right].
\end{equation}

\subsection{Non-oscillatory regime}
\label{app:non-oscillatory-regime}

We now consider the non-oscillatory regime $
1-\frac{m^2}{8}\geq0$.
For $l=5$, the corresponding equations are obtained from Eqs.~\eqref{x', y', z'} and
read
\begin{equation}\label{app:x-y-z-nonosc}
\left\{
\begin{array}{l}
x' = z, \\[6pt]
y' = -3\rho^2 y^2 + \dfrac{z^2 + E^2}{\rho^2}
+ \dfrac{12\lambda}{6-m}\rho^{2}\cos(\varphi + mx), \\[8pt]
z' = -2\dfrac{\rho'}{\rho}z
-\dfrac{2\lambda m}{6-m}\rho^{4}\sin(\varphi + mx).
\end{array}
\right.
\end{equation}

Introducing the variable
\begin{equation}
f\equiv\frac{z}{\rho^2},
\end{equation}
the system can be rewritten as
\begin{equation}
\left\{
\begin{array}{l}
f\dfrac{dy}{dx}
=
-3y^2+f^2+\dfrac{E^2}{\rho^4}
+\dfrac{12\lambda}{6-m}\cos(\varphi+mx),
\\[10pt]
f\dfrac{df}{dx}
=
-4yf
-\dfrac{2\lambda m}{6-m}\sin(\varphi+mx),
\end{array}
\right.
\end{equation}
where we have assumed $f\neq0$. Expanding around the fixed point as
\begin{equation}
x=\bar{x}_n+\xi,
\qquad |\xi|\ll1,
\end{equation}
and using Eq.~\eqref{y^bar^2}, the system becomes
\begin{equation}\label{app:perturbative-f-y}
\left\{
\begin{array}{l}
f\dfrac{dy}{d\xi}
=
-3y^2
+f^2
+\dfrac{E^2}{\rho^4}
+3\bar y^2\left(1-\dfrac12m^2\xi^2\right),
\\[10pt]
f\dfrac{df}{d\xi}
=
-4yf
-\dfrac12\bar y^2m^2\xi.
\end{array}
\right.
\end{equation}

Using the asymptotic solutions in Eqs.~\eqref{xi(rho)} and
\eqref{zeta(rho)}, one finds
\begin{equation}
f(\xi)
=
-2\bar y(1-\mu)\xi,
\label{app:f-xi}
\end{equation}
where $\mu\in[0,1)$. Notice that the contribution proportional to
$E^2/\rho^4$ is subleading with respect to $\xi^2$ in the asymptotic regime
and can therefore be neglected at leading order.

Motivated by the asymptotic behaviour of the solutions, we introduce the
ansatz
\begin{equation}
y(\xi)
=
\bar y\left(1+y^*m^2\xi^2\right),
\label{app:y-ansatz-nonosc}
\end{equation}
which, substituted into Eq.~\eqref{app:perturbative-f-y}, gives
\begin{equation}\label{app:y-star}
y^*
=
-\frac{1}{2(1+\mu)}.
\end{equation}

The corresponding correction to the equation of state parameter is
\begin{equation}
\delta w(\rho)
=
-\frac{y'}{\rho^2y^2}
=
4m^2y^*(1-\mu)\xi^2.
\label{app:deltaw-xi-nonosc}
\end{equation}
Using the asymptotic solutions for $\xi$ in Eq.~\eqref{xi(rho)}, this becomes
\begin{equation}\label{app:deltaw-nonosc}
\delta w(\rho)
=
\frac{\delta w_0}{\rho^4}
\begin{cases}
\rho^{4\mu},
&
0<\mu<1,
\\[8pt]
\log^2\rho,
&
\mu=0,
\end{cases}
\end{equation}

where

\begin{equation}
\delta w_0
\equiv
-16(1-\mu)^2B^2.
\end{equation}

\printbibliography

\end{document}